%% file: paper.tex
\documentclass[sigplan,nonacm]{acmart}

\usepackage{graphicx}
\usepackage{xcolor}
\usepackage{xspace}
\usepackage{subcaption}
\usepackage[font=small,labelfont=bf]{caption}
\usepackage{enumitem}
\usepackage{makecell}
\usepackage{microtype}
\usepackage{url}
\usepackage{seqsplit}
\usepackage{multirow}
\usepackage{enumitem}
\usepackage{xspace}
\usepackage{pifont}

\setitemize{noitemsep,parsep=0em,partopsep=0pt}

\newcommand{\para}[1]{\noindent{\bf #1}}
\newcommand\sysname{\textsc{TierCheck}\xspace}

\begin{document}

\title[]{TierCheck: Tiered Checkpointing for Fault Tolerance in Large Language Model Training}

\author{Shujie Han$^1$, Feng Jiang$^1$, Patrick P. C. Lee$^2$, Xiao Zhang$^1$,
Zhijie Huang$^1$,\\
Nannan Zhao$^1$, Xiaonan Zhao$^1$, and Lichen Pan$^3$ \\
$^1$Northwestern Polytechnical University \ \
$^2$The Chinese University of Hong Kong \\
$^3$National University of Defense Technology
}

\renewcommand{\shortauthors}{}

\input{abstract}

\maketitle

\begin{sloppypar}
\input{introduction}
\input{background}
\input{design}
\input{implementation}

\input{evaluation}
\input{related}
\input{conclusion}
\end{sloppypar}

\bibliographystyle{ACM-Reference-Format}
\bibliography{paper}


\end{document}

%% file: abstract.tex
\begin{abstract}
Large Language Model (LLM) training is frequently interrupted by a heterogeneous spectrum of failures, from common GPU crashes to catastrophic cluster-wide outages. Existing checkpointing systems rely on monolithic, single-tier storage backend, forcing a trade-off between state-saving overhead and recovery speed. We propose \sysname, a cluster-aware tiered checkpointing system that aligns storage placement with failure heterogeneity. \sysname adopts a three-tier design that maintains lightweight differential checkpoints in local and peer memory for fast localized recovery, while asynchronously migrating heavyweight base checkpoints to remote persistent storage. It also ensures strict global consistency across tiers without stalling training, and achieves fast cluster-aware checkpoint restoration during recovery. Evaluations on models up to 40~billion parameters show that \sysname achieves low training overhead, reduces end-to-end checkpointing time to under 10\,s, and supports high-frequency checkpointing, ultimately striking an optimal balance between low-overhead persistence and fast recovery.


\end{abstract}

%% file: introduction.tex
\section{Introduction}
\label{sec:intro}

In Large Language Model (LLM) training, failures are both prevalent and diverse.  As model sizes scale to trillions of parameters \cite{achiam23,grattafiori24}, training requires thousands of GPUs running continuously for months \cite{jiang24,scao22bloom}, turning unexpected failures into routine daily occurrences. For instance, during a 54-day period of Llama-3 training at Meta \cite{grattafiori24}, 419 of 466 job interruptions are unexpected failures at GPU and node levels, of which GPU failures alone account for 58.7\%. Modern data centers also experience rack failures where multiple nodes within the same rack are unavailable simultaneously, albeit rarely \cite{zhang19,ford10}. This skewed statistical distribution reveals the {\em heterogeneity} of failures in large-scale LLM training. 

To mitigate the loss of training progress, {\em periodic checkpointing} remains the standard fault-tolerance mechanism \cite{elnozahy02}.  In LLM training, massive training states (e.g., model weights and optimizer states) are intricately partitioned across thousands of GPUs via multi-dimensional parallelisms (e.g., Tensor Parallelism \cite{shoeybi19}, Pipeline Parallelism \cite{huang19}, and ZeRO Data Parallelism \cite{rajbhandari20zero}).  By regularly saving distributed training states, a training job can resume from the most recent globally consistent state after a failure. Recent work has extensively optimized checkpointing along three axes: (i) {\em checkpointing overhead}, which minimizes I/O blocking overhead to maximize training throughput via asynchronous persistence \cite{mohan21,maurya24,huang25}; (ii) {\em checkpoint size reduction}, which compresses states \cite{yao25} or exploits redundancy elimination \cite{liu26} to enable higher checkpointing frequencies; and (iii) {\em recovery efficiency}, which leverages host CPU memory to rapidly restore training states and minimize downtime \cite{wang23}.

However, existing checkpointing systems apply an {\em all-or-nothing} persistence strategy that largely ignores the heterogeneity of failures. In-memory systems (e.g., Gemini \cite{wang23}) prioritize fast recovery by saving snapshots in CPU memory, but are vulnerable to node crashes. Conversely, systems that write to local or remote persistent storage (e.g., CheckFreq \cite{mohan21}) offer stronger fault tolerance but suffer from severe I/O bottlenecks during recovery. Even advanced techniques that reduce checkpoint sizes via incremental saving (e.g., \cite{yao25,liu26}) are still bottlenecked either by the computational cost of sequential state replay \cite{yao25} or the overhead of redundant state detection \cite{liu26} during checkpoint restoration. With the tight binding of a monolithic storage backend to diverse failure modes, these systems fail to strike an optimal balance between low-overhead persistence and fast recovery.

To address this gap, an intuitive approach is to decouple checkpoint persistence across a hierarchy of storage layers. 
However, realizing a multi-tiered architecture introduces three critical system-level challenges spanning the entire checkpoint lifecycle:

\begin{itemize}[leftmargin=*]
\item
{\bf Checkpoint saving:} Offloading massive training states from GPUs competes for system resources. The challenge is to efficiently capture and route these states to appropriate tiers without stalling ongoing training.  
\item
{\bf Checkpoint retrieval:}  Because failures vary in frequency and severity, the recovery path must be optimized for the most common incidents (e.g., GPU and node failures) while still guaranteeing a robust fallback for rare, catastrophic events (e.g., rack-level power outages).
\item
{\bf Checkpoint reclamation:}  Extreme bandwidth disparities between tiers inevitably cause slower storage to lag behind fast volatile memory. The system must determine precisely when it is safe to garbage-collect volatile checkpoints across the cluster.
\end{itemize}

We design \sysname, a cluster-aware tiered checkpointing system aiming to tolerate heterogeneous failures in LLM training.  \sysname distributes checkpoint data across three tiers: {\em Tier-1 (local volatile memory)}, which retains local checkpoints for fast recovery, {\em Tier-2 (neighbor volatile memory)}, which asynchronously replicates checkpoints to a peer node for tolerating single-node failures, and {\em Tier-3 (remote persistent storage)}, which durably migrates checkpoints to a remote backend for tolerating rack failures. 

To overcome the challenges of multi-tier checkpointing, \sysname optimizes the
checkpoint lifecycle across three key phases. First, for {\em checkpoint
saving}, \sysname decouples the persistence of training states. It frequently
captures and compresses incremental updates into fast volatile memory (Tier-1
and Tier-2) to prevent bursty I/O from stalling training, while periodically
buffering complete states at a much lower frequency and asynchronously migrating
them to remote storage in the background.  Second, for {\em checkpoint
retrieval}, \sysname dynamically routes the restoration path. It prioritizes
fast recovery from Tier-1 and Tier-2 to quickly resolve common incidents (e.g.,
software and node failures), while seamlessly falling back to
remote storage (Tier-3) for catastrophic rack failures. Third, for {\em checkpoint
reclamation}, \sysname safely governs the lifecycle of historical states. It
strictly purges volatile replicas only after the corresponding base checkpoint
is durably committed to remote storage, guaranteeing global consistency without
risking data loss.

We implement \sysname atop DeepSpeed~\cite{rasley20}, achieving seamless
compatibility with 3D parallelism strategies, including Tensor
Parallelism~\cite{shoeybi19}, Pipeline Parallelism~\cite{huang19}, and ZeRO Data
Parallelism~\cite{rajbhandari20zero}. Our implementation natively supports the
entire checkpoint lifecycle, systematically orchestrating asynchronous saving, a
highly optimized retrieval path, and watermark-driven reclamation.
We evaluate \sysname on a testbed of 16 NVIDIA A800 (80\,GiB) GPUs provisioned
from a distributed cloud platform. Experimental results demonstrate that
\sysname reduces the checkpointing overhead by 62.8-82.7\% compared to
state-of-the-art systems (e.g., CheckFreq \cite{mohan21} and Gemini
\cite{wang23}). Furthermore, \sysname achieves a 3.6-6.9$\times$ recovery
speedup for software failures and 1.8-3.5$\times$ for node failures. Even under
worst-case rack failures, it remains 1.3-2.5$\times$ faster than baselines, all
without compromising global model convergence.  We will open-source the \sysname
prototype in the final paper.

%% file: background.tex
\section{Background and Motivation}
\label{sec:background}


\subsection{LLM Training}
\label{subsec:background_basics}

\para{Basics.} The LLM training process is an iterative optimization loop. Each training iteration comprises a {\em forward pass} that computes the loss over a data batch and a {\em backward pass} that computes the gradients. The {\em optimizer} (e.g., Adam \cite{kingma15}) uses these gradients along with its internal states (e.g., first and second moments) to update model weights.  

\para{Training state size.} Checkpointing overhead scales directly with training state size. Modern mixed-precision training creates an asymmetry between the lightweight representation used for computation and the heavier training state that must be checkpointed for exact recovery. Let $\Phi$ denote the total number of model parameters.  In mixed-precision training, model weights are stored in 16-bit half-precision floating-point format (FP16), consuming 2$\Phi$ bytes.  To prevent numerical underflow, the Adam optimizer additionally maintains a 32-bit full-precision floating-point (FP32) master copy of the weights alongside its first and second moment tensors, which together consume 12$\Phi$ bytes.  Thus, a full training snapshot requires 14$\Phi$ bytes in total, dominating both memory and storage footprints at scale.

\para{Parallel mechanisms.} To accommodate trillion-parameter LLMs, modern
training frameworks combine multiple parallelism strategies.  {\em Data
Parallelism (DP)} replicates the model to process independent data batches
concurrently, often augmented by the {\em Zero Redundancy Optimizer (ZeRO)}
\cite{rajbhandari20zero}. ZeRO progressively shards training states across DP
{\em ranks} (i.e., individual GPU workers) to minimize memory footprints: ZeRO-1
shards optimizer states, ZeRO-2 adds gradient sharding, and ZeRO-3 further
shards model weights.  {\em Tensor Parallelism (TP)} \cite{shoeybi19} partitions
individual parameter matrices across GPUs, while {\em Pipeline Parallelism (PP)}
\cite{huang19} assigns consecutive model layers to different nodes for
concurrent micro-batch processing. 


\subsection{Checkpointing}
\label{subsec:motivation}

\para{Training failures.} Failures in large-scale LLM training are heterogeneous in both frequency and severity. We categorize training failures into three classes: (i) {\em Software failures} are software-triggered incidents such as CUDA kernel crashes, runtime bugs, or malformed input data \cite{zhang20}.  Since the underlying hardware remains healthy, the process can simply be restarted on the same node, preserving local host memory contents.  (ii) {\em Node failures} are host- or device-level hardware crashes caused by CPU, DRAM, motherboard, attached GPU devices, or local power faults, and dominate hardware-related incidents in production clusters \cite{grattafiori24}. Since the affected hardware becomes unavailable, recovery requires replacing the failed component or node, and all volatile state on that hardware is lost. (iii) {\em Rack failures}, such as rack-level power outages, top-of-rack switch failures, or cooling faults, simultaneously affect multiple nodes within the same rack. They are rare but catastrophic: they wipe out entire failure domains, including all local and peer volatile memory, leaving only remote durable storage as a recovery source.

\begin{table*}[t]
\setlength{\tabcolsep}{4pt}
\renewcommand{\arraystretch}{1.1}
\centering
\small
\caption{Comparison of checkpointing systems and their recovery mechanisms under heterogeneous failures. $\checkmark$ indicates efficient and exact recovery; $\sim$ indicates recovery with significant trade-offs (e.g., slow performance, approximation, or massive hardware redundancy); $\times$ indicates inability to recover without catastrophic fallback or failure.}
\label{tab:failure_handling}
\vspace{-9pt}
\begin{tabular}{l|l|l|l}
\hline
\makecell[c]{\textbf{Systems}} & \makecell[c]{\textbf{Software failures}} & \makecell[c]{\textbf{Node failures}} &  \makecell[c]{\textbf{Rack failures}} \\
\hline
\hline
\makecell[l]{CheckFreq \cite{mohan21}, DataStates\\-LLM \cite{maurya24}, PCcheck \cite{strati25},\\ AsymCheck \cite{ming26}} & \multicolumn{3}{c}{$\sim$ Full remote checkpoint fetch (Bounded by slow persistent storage I/O)} \\
\hline
LowDiff \cite{yao25} & \multicolumn{3}{c}{$\sim$ Remote compressed fetch + Sequential differential replay} \\
\hline
AdaCheck \cite{liu26} & \multicolumn{3}{c}{$\sim$ Remote half-precision fetch + Decompression overhead} \\
\hline
Gemini \cite{wang23} & $\checkmark$ Local node fetch & $\checkmark$ Peer node fetch & $\times$ Volatile replicas wiped out \\
\hline
CheckFree \cite{blagoev26} & $\sim$ Pipeline averaging & $\sim$ Pipeline averaging & $\times$ Not designed for multi-node failures \\
\hline
Swift \cite{zhong24} & $\checkmark$ DP group replicas & $\checkmark$ DP group replicas & $\times$ Hardware redundancy overwhelmed \\
\hline
FT-HSDP \cite{salpekar26} & $\sim$ Drop failed replica & $\sim$ Drop failed replica & $\sim$ Remote fetch (Extremely costly) \\
\hline
{\bf \sysname (Ours)} & $\checkmark$ \textbf{Tier-1 Local memory} & $\checkmark$ \textbf{Tier-2 Peer memory} & $\checkmark$ \textbf{Tier-3 Base anchor + Differential replay} \\
\hline
\end{tabular}
\end{table*}

\para{Checkpointing classification.} LLM training systems tolerate failures by periodically saving training states to {\em remote persistent storage} (e.g., Lustre \cite{schwan03}, Ceph \cite{weil06}, Amazon S3 \cite{bornholt21}), which serves as the ultimate recovery source. Existing checkpointing mechanisms differ along two dimensions. By {\em placement}, checkpoints are either {\em in-memory} (i.e., local or peer volatile memory for low-latency recovery) or {\em persistent} (i.e., remote durable backends for strong failure isolation). By {\em format}, checkpoints are either {\em full} (capturing the complete state for direct, exact restoration) or {\em differential} (recording only the incremental updates from training iterations to reduce checkpoint volume). Full checkpoints enable $O(1)$ recovery but incur severe I/O overhead; differential checkpoints greatly reduce save costs, but recovery requires sequentially replaying an update chain from the most recent full checkpoint, incurring $O(L)$ replay cost proportional to the chain length $L$. While $L$ is typically bounded to tens to hundreds by the base interval, replaying even a short sequence requires repeated computation and communication over the massive training state, making the cumulative recovery time highly non-trivial. 

Because training states are partitioned across ranks, checkpoints are materialized as per-rank {\em shards} rather than single globally visible files. In this paper, a {\em base checkpoint} denotes a single rank's shard of a full snapshot, collectively serving as the {\em recovery anchor} (i.e., the start point for differential checkpoint replay). A {\em differential checkpoint} denotes one rank's shard of the incremental updates. The number of iterations between consecutive base checkpoints defines the \emph{base checkpoint interval}.

\para{Compressing differential checkpoints.} Differential checkpoints are further compressed as raw updates are too large for high-frequency saving. One approach is \textit{sparse compression} (e.g., Top-K selection), which extracts only a small subset of large-magnitude entries rather than storing the full dense update \cite{aji17,yao25}. While effective for large tensors with heavily concentrated updates, sparse compression incurs index metadata overhead that can outweigh its spatial benefits on smaller tensors \cite{renggli19}.  An alternative approach is {\em quantization-based compression} \cite{alistarh17qsgd}, which maps high-precision floating-point numbers to a limited set of discrete levels (e.g., 8-bit integers (INT8)). This preserves the dense tensor layout while uniformly reducing the bit width of each element, offering a more efficient trade-off for small parameter vectors. 

Existing checkpointing systems often optimize for a single placement tier, a single checkpoint format, or a single dominant failure class, as shown in Table~\ref{tab:failure_handling}. They excel for a subset of failures but underperform or fail for others.

\para{Limitations of single-tier checkpointing.}  We define a {\em tier} as a storage backend characterized by a specific durability model and bandwidth/latency profile (e.g., local volatile memory, peer volatile memory, or remote persistent storage). No single tier can simultaneously satisfy the requirements imposed by all three failure classes. A remote-only design provides durable recovery during catastrophic outages, but forces even the most common software and single-node failures to incur full remote-storage fetch latency. A volatile-only design enables fast, localized restoration for common failures, but offers no protection once the failure destroys the hosting failure domain, as in node or rack failures. A differential-heavy design reduces checkpoint volume, but the recovery cost grows linearly with replay length and is limited by the availability of a recent full checkpoint. This mismatch is most apparent through the lens of {\em failure survivability}: after a software failure, local memory states remain available; after a node failure, peer memory states still survive; after a rack failure, only remote durable storage can be leveraged. Since the cheapest surviving recovery source changes with failure scope, no single-tier strategy can simultaneously provide locality, durability, and low overhead.

\para{The case for tiered checkpointing.} Instead of coupling all checkpoints to a single backend, checkpoint placement should be aligned with failure domains such that each tier handles the failure scenarios it can optimally serve. A tiered architecture allows (i) fast local volatile storage to absorb frequent software failures, (ii) peer memory to recover from node failures without remote I/O, and (iii) a remote durable tier to recover from rack failures. This separation of fast persistence, localized recovery, and durable fallback preserves low steady-state overhead while guaranteeing exact recovery across the full failure spectrum. 

\subsection{Challenges of Tiered Checkpointing}
\label{subsec:challenges}

Realizing a tiered checkpointing architecture in large-scale LLM training clusters introduces three challenges.

\para{Checkpoint saving.} A tiered system must decide how to distribute checkpoint data across local memory, peer memory, and remote storage without degenerating into naive three-way replication, which amplifies storage and bandwidth overhead and stalls the training loop behind massive checkpoint I/O. This is compounded by the deep sharding of model weights, gradients, and optimizer states across TP, PP, DP, and ZeRO partitions, making the materialization of a full snapshot inherently expensive. The challenge is thus to construct tier-specific checkpoint contents of appropriate granularity and schedule their transfers asynchronously, so that each tier stores only the data it needs while keeping the critical training path short.

\para{Checkpoint retrieval.} Under heterogeneous failures, the system must quickly identify and exploit the cheapest surviving recovery path rather than blindly restoring from a fixed backend. The optimal path depends on which tiers remain intact: a software failure may be served entirely by local memory, a node failure by peer memory, and a rack failure by the remote persistent tier. Moreover, the total recovery cost is determined not only by data retrieval latency but also by the amount of replay or recomputation required after loading. The challenge is to jointly minimize pull time and re-execution time while ensuring seamless, exact fallback when faster tiers are unavailable, stale, or incomplete.

\para{Checkpoint reclamation.} Since remote persistence is orders of magnitude slower than fast-tier writes, recovery should retrieve a local or peer memory copy whenever possible. However, retaining checkpoints in local and peer memory indefinitely is infeasible, as these volatile resources are shared with the training process and will quickly become a bottleneck. The challenge is to determine precisely when a checkpoint becomes globally safe to reclaim, given the asynchronous and heterogeneous completion times across all tiers, and how to perform coordinated garbage collection across the cluster without creating data gaps.

%% file: design.tex
\begin{figure*}[t]
\centering
\begin{tabular}{c}
\includegraphics[width=6.6in]{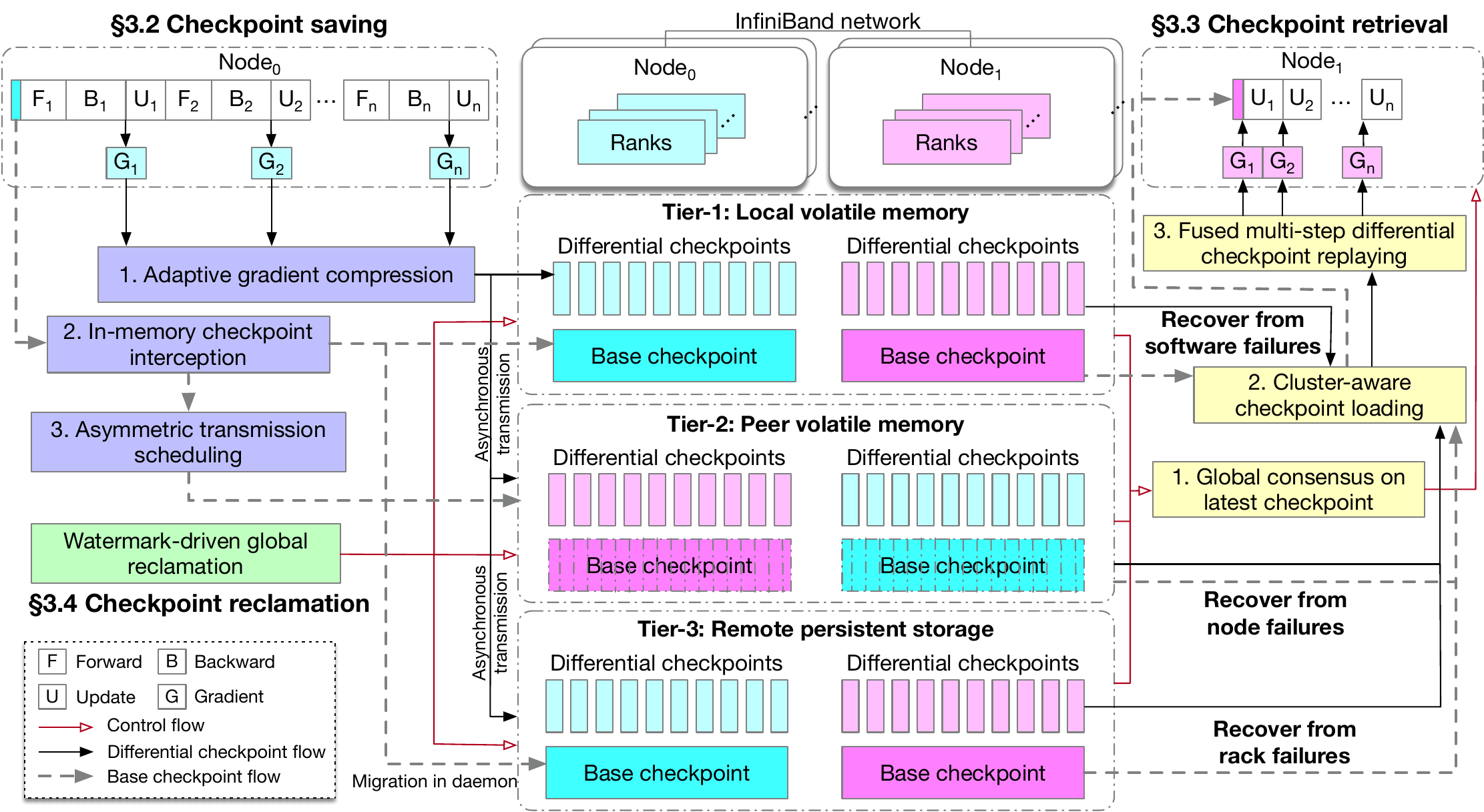} 
\end{tabular}
\vspace{-9pt}
\caption{Architectural overview of \sysname.}
\label{fig:architecture}
\vspace{-6pt}
\end{figure*}

\section{Design of \sysname}
\label{sec:design}

\sysname is a cluster-aware tiered checkpointing system designed for tolerating heterogeneous failures in LLM training. The design follows a simple principle: separating checkpoint contents by size, production frequency, and failure coverage, then binding each class of data to the storage tier that serves it most efficiently. We first present the overall architecture (\S\ref{subsec:overview}), and then describe how \sysname saves (\S\ref{subsec:save}), retrieves (\S\ref{subsec:retrieve}), and reclaims (\S\ref{subsec:reclaim}) checkpoints.

\subsection{Architectural Overview}
\label{subsec:overview}

Figure~\ref{fig:architecture} shows the architectural overview of \sysname.  We consider a training cluster composed of multiple machines, each equipped with multiple GPUs and interconnected through InfiniBand NICs. To align storage with heterogeneous failures (\S\ref{subsec:motivation}), \sysname organizes checkpoint data into three tiers. \textbf{Tier-1} is local volatile memory on the same node, used for the fastest checkpoint writes and local recovery. \textbf{Tier-2} is peer memory on a physically adjacent node, used to survive single-node failures. Here, a {\em peer} is defined physically rather than by logical DP/TP/PP groups: each rank (as defined in \S\ref{subsec:background_basics}) backs up its checkpoint to the corresponding rank on an adjacent machine through a node-level ring mapping, so replicas leave the local node's failure domain. In particular, a Tier-2 peer intentionally resides in the {\em same rack} as its Tier-1 node. This design is a deliberate, failure-frequency-aware trade-off: rack failures are exceedingly rare, whereas single-node failures are the dominant hardware-level incident class \cite{dean09,ford10,zhang19}. Intra-rack peer placement thus eliminates the latency and bandwidth costs of cross-rack network transfers, which would otherwise slow the high-frequency Tier-2 replication path, at the expense of protection against the rare rack failures. Crucially, this residual risk is fully absorbed by \textbf{Tier-3}, a remote persistent storage backend that serves as the durable fallback for rack failures of any scope, including rack-level events. This cluster-aware design optimizes for the common case without sacrificing correctness for the rare case.

On top of these tiers, \sysname saves training state through two independent streams rather than naive three-way replication. The first stream contains high-frequency {\em differential} checkpoints, i.e., per-rank incremental updates that remain on the fast volatile path for fast recovery. The second stream contains low-frequency {\em base} checkpoints that periodically refresh the full-checkpoint recovery anchor.  These base checkpoints are migrated asynchronously toward more durable tiers.  This asymmetric organization preserves low overhead on the critical training path and still maintains an exact recovery anchor.

\sysname builds on three cooperating modules:
\begin{itemize}[leftmargin=*]
\item 
\textbf{Checkpoint saving.} This module addresses how checkpoints are created and placed across tiers. Its key techniques include: (i) {\em adaptive gradient compression}, which shrinks high-frequency differentials before they enter the I/O path; (ii) {\em in-memory checkpoint interception}, which intercepts the optimizer's internal serialization path to extract checkpoint data as reusable in-memory payloads without reconstructing the full state; and (iii) {\em asymmetric transmission scheduling}, which applies distinct batching and chunking strategies to the two checkpoint streams to avoid bursty interference with training communication.
\item 
\textbf{Checkpoint retrieval.} This module decides how \sysname resumes from the most efficient surviving tier after a failure. Its key techniques include: (i) {\em global consensus on latest checkpoint}, which identifies the latest globally recoverable version across ranks; (ii) {\em cluster-aware checkpoint loading}, which maps checkpoint shards to the current cluster topology and avoids unnecessary data movement; and (iii) {\em fused multi-step differential checkpoint replaying}, which reconstructs the latest state by replaying a bounded sequence of differential checkpoints efficiently.
\item 
\textbf{Checkpoint reclamation.} This module decides when old volatile checkpoints can be safely removed. Its key mechanism is {\em watermark-driven global reclamation}, in which Tier-1 and Tier-2 histories are reclaimed only after the corresponding Tier-3 base checkpoint is globally safe.  
\end{itemize}

\subsection{Checkpoint Saving}
\label{subsec:save}

\para{Design goal.} The saving path must resolve the tension that frequent saves are necessary to reduce rollback, but pushing the full training state through tiers on every save incurs significant overhead. \sysname separates the save path into two asymmetric streams. A lightweight differential stream is generated frequently, exposed first on Tier-1 and Tier-2 for fast recovery, and transferred asynchronously to Tier-3, where explicit completion tracking ensures the checkpoint is fully committed to ensure broader failure coverage. A heavyweight base-checkpoint stream is generated less often and drained away from the critical training loop in the background. The key design goal is to generate each stream once, place it on the tiers that need it, and keep protection off the critical training path.

\para{Adaptive gradient compression.} Since precise optimizer differentials are complex to track, \sysname leverages gradients to form the high-frequency stream \cite{yao25,liu26}. By intercepting the optimizer-visible partitions already exposed in ZeRO training \cite{rajbhandari20zero}, \sysname captures and compresses these gradients before they enter the I/O pipeline. The compression strategy adapts to tensor size. For small tensors (e.g., fewer than $100$K elements), where sparse compression offers little space reduction (\S\ref{subsec:motivation}), \sysname applies dense INT8 quantization \cite{alistarh17qsgd}. For large tensors, \sysname targets a sparsification ratio of $k=0.01$ \cite{yao25}. Conventional Top-K selection is prohibitively expensive due to multiple global-memory passes \cite{aji17}.  Instead, \sysname employs a threshold-based sparse kernel: it estimates a magnitude threshold via sampling, then uses a single fused pass to extract and compact surviving entries into FP16 values and INT32 indices. To prevent index overflows in massive models, where a single tensor may contain billions of elements that exceed the INT32 range, oversized tensors are chunked before compression and safely rebased during reassembly. Ultimately, this adaptive design minimizes both the differential checkpoint size and the computational overhead of compression. 

\para{In-memory checkpoint interception.} The standard training framework's base checkpointing path is suboptimal for tiered storage. For example, DeepSpeed \cite{rasley20} treats persistence as a synchronous, terminal operation, writing serialized checkpoints directly to storage. This opaque design forces a tiered system to incur redundant disk I/O, i.e., writing the full checkpoint only to immediately read it back for peer transmission. To eliminate this write-then-read penalty, \sysname intercepts the internal \texttt{torch.save} path, capturing serialized base checkpoints directly as in-memory byte payloads.  This zero-copy interception allows a single reconstructed state to be simultaneously dispatched to local memory (Tier-1), peer replication (Tier-2), and background remote migration (Tier-3), completely isolating the critical training path from synchronous storage stalls.

\begin{figure}[!t]
\centering
\begin{tabular}{c}
\includegraphics[width=3.2in]{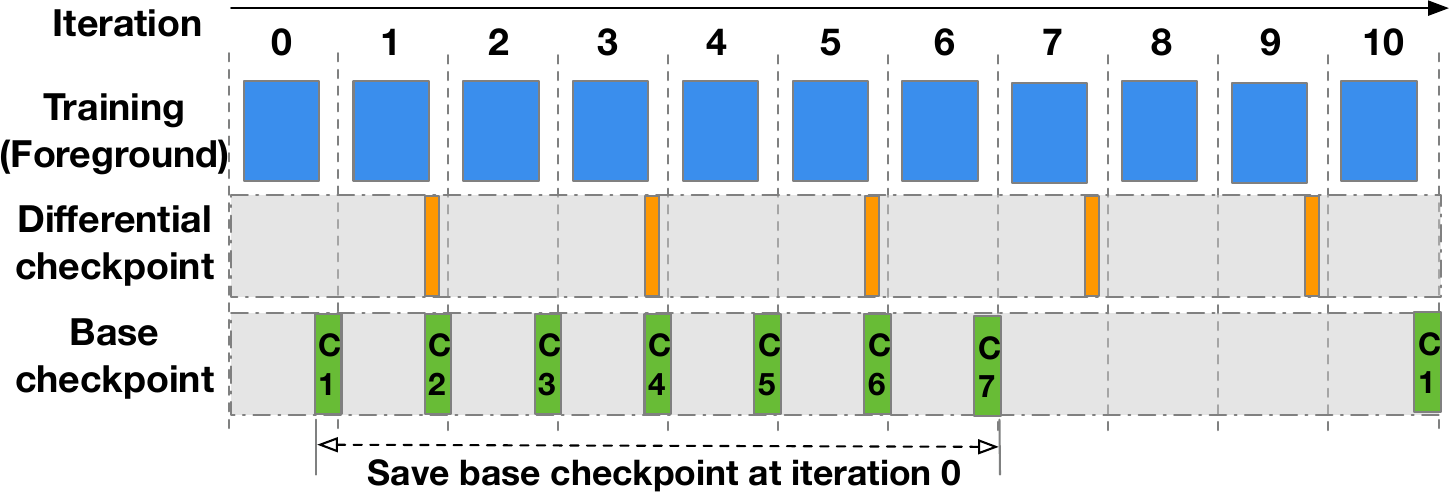} 
\end{tabular}
\vspace{-9pt}
\caption{Adaptive cross-tier transmission without stalling foreground training.}
\label{fig:transmission}
\vspace{-6pt}
\end{figure}

\para{Asymmetric transmission scheduling.} Moving state across tiers without stalling training is challenging. As shown in Figure~\ref{fig:transmission}, \sysname employs distinct strategies for the two checkpoint streams. For lightweight differential checkpoints (typically megabytes in size), \sysname batches multiple iterations into a compact byte stream using a configurable batching length $N$, where one batch denotes the differential checkpoints from $N$ consecutive iterations grouped together for batched saving and replaying \cite{yao25} (e.g., $N=2$ in Figure~\ref{fig:transmission}). Conversely, monolithic replication of gigabyte-scale base checkpoints would severely block training. To mitigate this, \sysname applies chunk-based transmission exclusively to Tier-2 base replication, dividing each base checkpoint into paced micro-chunks under a bounded per-iteration bandwidth budget (e.g., completing over 7 iterations as shown in Figure~\ref{fig:transmission}). The peer target is deterministically selected via a physical cross-node ring topology, mapping each rank directly to its counterpart on an adjacent machine. This one-to-one mapping ensures that replicas escape the local failure domain without centralized placement logic.

To keep peer protection from turning into a blocking transfer, replication must finish before the next base checkpoint is generated. Peer ranks first exchange their serialized payload sizes to establish the maximum bilateral transfer volume. \sysname then evenly divides this volume across the available training iterations within the base checkpoint interval, reserving a brief safety margin. Consequently, the chunk size adapts dynamically: longer intervals yield finer, less intrusive chunks, while tighter intervals demand larger chunks.  To avoid foreground interference, \sysname caps the maximum chunk size (e.g., 256\,MiB). If the base checkpoint interval is too short to drain a massive payload under this cap, \sysname extends the background transfer schedule across more iterations. As a fallback, if this extended transfer spills over into the next scheduled base checkpoint generation, \sysname temporarily stalls the foreground training loop to synchronously flush the remaining micro-chunks, guaranteeing memory safety and strict replica consistency before a new checkpoint overwrites the buffer. To prevent such synchronous stalls from degrading training performance, the base checkpoint interval is configured to a sufficiently large value (e.g., 100 iterations), ensuring that even massive payloads can fully drain in the background. Ultimately, this strategy exploits the natural computational gap in the common case while providing a safety net for extreme edge cases.

\subsection{Checkpoint Retrieval}
\label{subsec:retrieve}

\para{Design goal.} The retrieval path must prioritize both recovery speed and state consistency. Upon failure, \sysname must identify the latest globally consistent state before resuming from the fastest surviving storage tier.  To quantify checkpoint freshness, \sysname defines a {\em checkpoint version} by its corresponding training iteration index. 
Its retrieval comprises three steps: (i) establishing this latest globally recoverable state, (ii) determining which surviving checkpoint files each rank should consume under the current TP/PP/DP/ZeRO placement, and (iii) replaying the remaining differential checkpoints with bounded overhead.

\para{Global consensus on latest checkpoint.} 
After a failure, \sysname's recovery re-establishes the communicator across the restarted job and derives global consensus directly from surviving data, so that the consensus protocol does not depend on messages from the crashed process; it runs among the surviving or replacement ranks of the relaunched job. Specifically, each rank first identifies the highest base checkpoint version recoverable from its available tiers (local, peer, or remote). The cluster then computes a global minimum over these local observations to select the latest universally accessible base checkpoint. \sysname applies this identical distributed reduction to differential checkpoints to establish the maximum valid replay boundary. This decentralized consensus is necessary because durable protection inherently lags behind rapid, volatile saves; ranks often observe locally fresh but globally unrecoverable versions. By aggregating local recoverability via global minima, \sysname safely filters out transient inconsistencies and deterministically identifies the optimal recovery point.

\begin{figure}[t]
\centering
\begin{tabular}{c}
\includegraphics[width=2.5in]{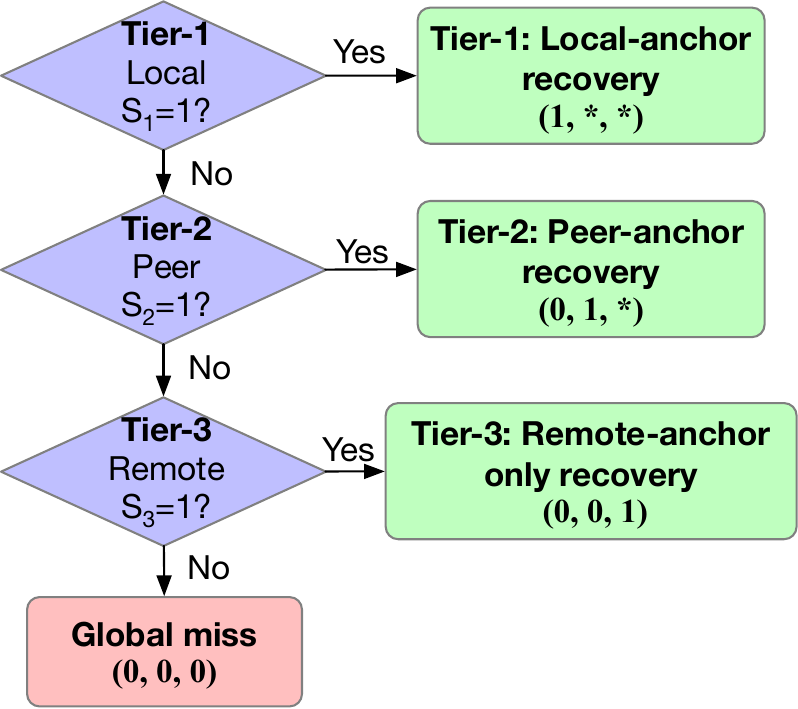} 
\end{tabular}
\vspace{-9pt}
\caption{Availability of base-checkpoint anchors across different storage tiers under heterogeneous failures.}
\label{fig:state}
\vspace{-6pt}
\end{figure}

To systematically categorize recovery scenarios, \sysname models the availability of a version's base checkpoint as a boolean tuple $(S_1, S_2, S_3) \in \{0,1\}^3$ (Figure \ref{fig:state}). Here, $S_1$ indicates whether the base checkpoint remains available in Tier-1 (local volatile memory), $S_2$ denotes its availability in Tier-2 (the designated peer replica), and $S_3$ signifies whether it has been durably committed to Tier-3 (remote persistent storage).  This availability tuple not only determines whether a specific version can serve as a valid recovery anchor, but also dictates the most cost-effective fallback source following a software failure, a node failure, or a rack failure.

\begin{itemize}[leftmargin=*]
\item 
\textbf{Local-anchor recovery ($S_1=1$).} This category covers states $(1,0,0)$, $(1,1,0)$, $(1,0,1)$, and $(1,1,1)$. When a software failure terminates training while the host node remains healthy, \sysname restores the base checkpoint directly from the local volatile copy, bypassing slower tiers entirely. The presence of additional peer or remote copies merely broadens the fallback space; it does not alter this optimal first choice.
\item 
\textbf{Peer-anchor recovery ($S_1=0, S_2=1$).} This category covers states $(0,1,0)$ and $(0,1,1)$. When a local node fails but its designated peer replica survives, \sysname reconstructs the missing base checkpoint from Tier-2. This volatile replica is significantly cheaper to access than a remote fetch. Even if a durable remote copy exists ($S_3=1$), Tier-3 acts strictly as a secondary backup rather than the preferred recovery path.  
\item 
\textbf{Remote-anchor-only recovery ($S_1=0, S_2=0, S_3=1$).} If both volatile copies are lost but the base checkpoint has been durably committed to Tier-3, \sysname falls back to this persistent remote base checkpoint. This scenario typically arises during rack failures and serves as the last-resort recovery path for any state that cannot be retrieved from the faster tiers.
\item 
\textbf{Global miss ($S_1=0, S_2=0, S_3=0$).} A catastrophic outage eradicates the latest base checkpoint across all tiers before a durable commit completes. Such events are exceedingly rare in practice, as they require a correlated cluster-wide failure to strike exactly during the narrow window of an asynchronous transfer to Tier-3. To safeguard against this, \sysname relies on a commit watermark to track durable completions; upon detecting an incomplete state, the system safely rolls back to the most recent base checkpoint version that is globally available and verified by the watermark.
\end{itemize}

The availability of differential checkpoints is evaluated independently of the base checkpoint availability tuple. Once the cluster reaches consensus on the latest recoverable base checkpoint version, each rank independently determines the most recent sequence of differential checkpoints available across its local, peer, and remote tiers. A subsequent global minimum operation then establishes the safe replay boundary for the entire cluster. In essence, the tuple $(S_1, S_2, S_3)$ dictates {\em whether} a version can serve as the exact recovery anchor, whereas the availability of differential checkpoints determines {\em how far} recovery can advance beyond that anchor. During the loading phase, differential checkpoints adhere to the same cost-ordered retrieval cascade: local memory first, followed by the designated peer, and finally Tier-3 for any remaining unavailable differential checkpoints.

\para{Cluster-aware checkpoint loading.} 
Once the target checkpoint version is chosen, \sysname needs to address training parallelism in checkpoint loading. 
Because modern training frameworks combine multiple parallelism strategies (TP/PP/DP/ZeRO), a recovered job cannot simply open a single shared checkpoint directory. Instead, each rank must selectively assemble the specific checkpoint shards dictated by its role in the active parallel layout and retrieve the associated metadata, such as version manifests and commit markers, to correctly map the distributed model and optimizer states to the target version.


To address this, \sysname performs {\em cluster-aware checkpoint loading}, following a tiered cascade ordered by retrieval cost. First, a recovering rank attempts to reuse its own local volatile checkpoints (Tier-1) if they survived the failure. If these are unavailable, the rank pulls the missing files directly from its assigned cross-node peer (Tier-2), mirroring the ring topology used during checkpoint saving. Next, for model states shared across multiple ranks under the active parallel layout, \sysname fetches them from surviving intra-cluster peers. Only the remaining unavailable files are fetched from Tier-3. Even during a durable fallback to remote storage, loading remains highly selective: a single leader process on each node enumerates exactly the files required by its local ranks and acts as a proxy to fetch only those specific shards, avoiding a full directory-wide remote download.

\begin{figure}[t]
\centering
\begin{tabular}{@{\ }c@{\hspace{2em}}c}
\hspace{-1em}
\includegraphics[width=1.68in]{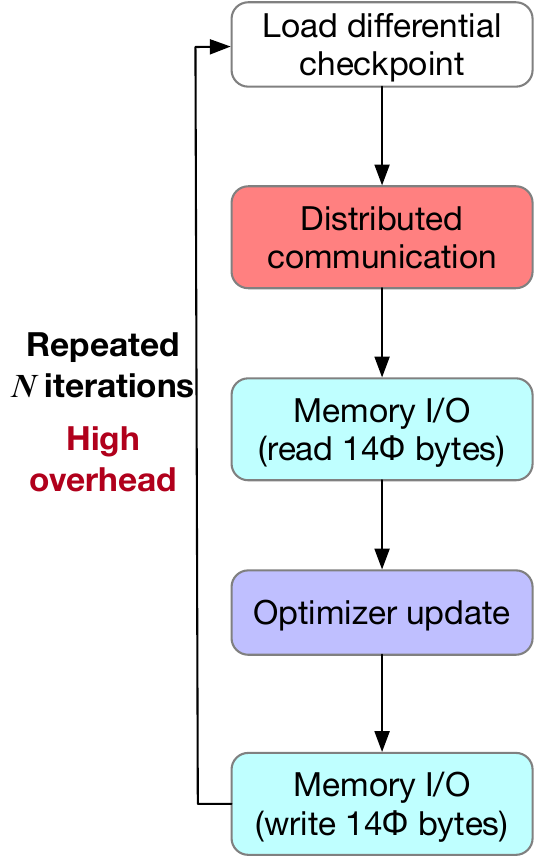} &
\includegraphics[width=1.35in]{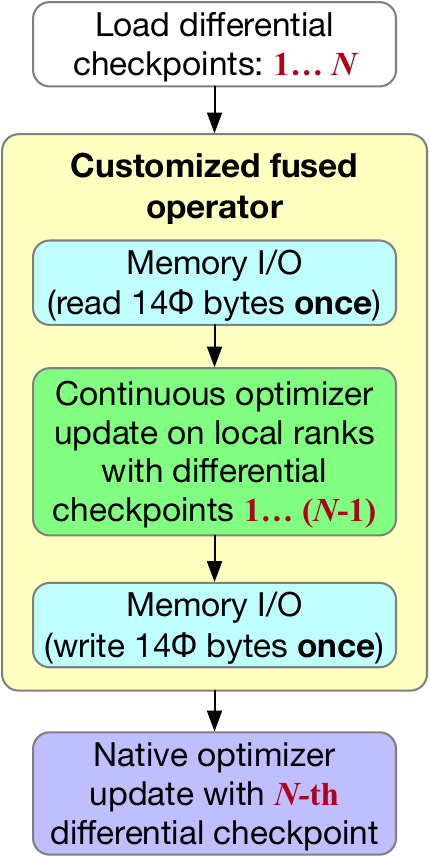} 
\\
\mbox{\small (a) Native replay} &
\mbox{\small (b) Fused multi-step replay}
\end{tabular}
\vspace{-9pt}
\caption{Comparison of native and fused multi-step differential checkpoint replaying.}
\label{fig:replay}
\vspace{-6pt}
\end{figure}

\para{Fused multi-step differential checkpoint replaying.} Even with efficient cluster-aware loading, differential checkpoint replay remains on the critical recovery path. Because the Adam optimizer is stateful, iterations cannot simply be accumulated or parallelized. Each iteration rigorously depends on the first and second moments produced by all preceding iterations. Consequently, a naive recovery must invoke the standard distributed optimization routine for every replayed iteration. As shown in Figure~\ref{fig:replay}(a), this naive approach incurs massive penalties at every iteration: it needlessly triggers expensive distributed communications (e.g., cluster-wide collective operations to synchronize sharded gradients), repeatedly reads and writes the entire $14\Phi$-byte training state to GPU memory, and suffers from severe CPU-to-GPU dispatch overhead caused by launching numerous fine-grained, per-parameter operators from the native Python runtime.

To eliminate these overheads, \sysname introduces {\em fused multi-step differential checkpoint replaying}. Rather than replaying the entire historical chain in a single monolithic pass, \sysname streams the recovered differential checkpoints and replays them in bounded batches. Each replay batch contains $N$ consecutive differential checkpoints (\S\ref{subsec:save}). As shown in Figure~\ref{fig:replay}(b), all $N$ checkpoints in a batch are first loaded onto the recovery path. A customized fused operator then consumes the first $N-1$ checkpoints together: it reads the model weights, first moments, and second moments exactly once, applies the corresponding $N-1$ incremental gradients in temporal order locally on each rank, and writes the final results back to memory. Because these updates are applied directly to the local parameter partitions on each rank, this design seamlessly supports ZeRO-1/2/3 without requiring intermediate cross-rank synchronizations.  While minor floating-point variations inherent to fused kernels and gradient compression preclude strict bitwise exactness, \sysname maintains strong algorithmic equivalence to the baseline optimizer, paying communication costs exactly once per batch to accelerate replay by orders of magnitude.

\subsection{Checkpoint Reclamation}
\label{subsec:reclaim}

\para{Design goal.} The reclamation path must resolve a tension in tiered checkpointing: volatile checkpoints cannot be retained indefinitely due to Tier-1 and Tier-2 memory constraints, yet they cannot be prematurely discarded based solely on local freshness. As checkpoint saving proceeds asynchronously, a newly generated base checkpoint may manifest in the volatile tiers of a fast node while slower nodes are still processing it and remote commits remain in flight. Consequently, a naive local-age policy that eagerly reclaims historical differential checkpoints preceding the newest locally visible state is incorrect under heterogeneous failures, as local recency provides no guarantee of cluster-wide redundancy. Reclamation must thus be governed by global recoverability rather than per-node progress.

\begin{figure}[t]
\centering
\begin{tabular}{@{\ }c@{\ }c}
\multicolumn{2}{c}{\includegraphics[width=3in]{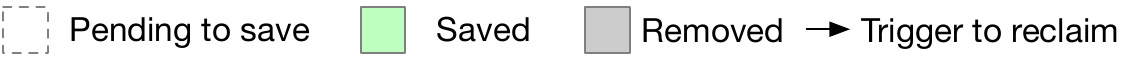}} \\
\includegraphics[width=1.5in]{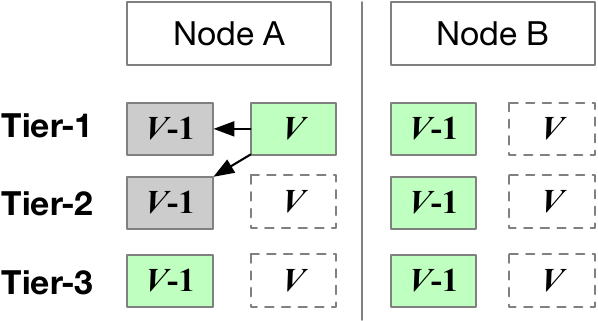} &
\includegraphics[width=1.5in]{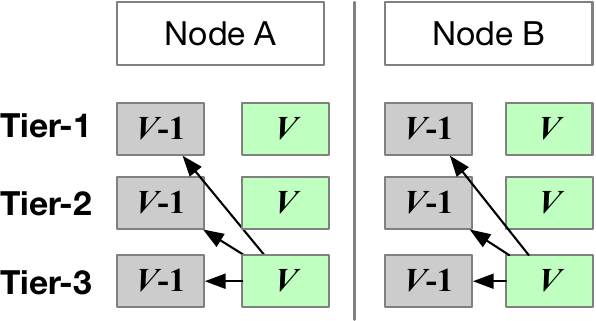} 
\\
\mbox{\small (a) Naive} &
\mbox{\small (b) Watermark-driven}
\end{tabular}
\vspace{-9pt}
\caption{Comparison of naive and watermark-driven global reclamation.}
\label{fig:reclaim}
\vspace{-6pt}
\end{figure}

\para{Watermark-driven global reclamation.} \sysname resolves this synchronization challenge by deferring garbage collection using a strictly monotonic global watermark. This watermark tracks the highest checkpoint version whose base checkpoint is fully persisted to Tier-3, thereby establishing a globally stable replay anchor. To illustrate the necessity of this mechanism, consider the naive reclamation in Figure~\ref{fig:reclaim}(a). A fast node (Node A) locally completes base checkpoint $V$ and eagerly reclaims the volatile recovery chain rooted at $V-1$. This premature deletion includes its Tier-1 and Tier-2 base copies of $V-1$, as well as any subsequent differential checkpoints. Meanwhile, a slower node (Node B) still holds $V-1$ in its volatile tiers. Because transferring massive model states takes significant time, the new base checkpoint $V$ may still be uploading to the slow Tier-3 storage, leaving $V-1$ as the only complete, durably committed remote anchor. 

If Node A fails, its local Tier-1 memory is lost, and the system must fall back to the remote anchor at $V-1$. However, the sequence of differential checkpoints bridging $V-1$ and $V$ is now fragmented. While some incremental updates may have already reached Tier-3 or survived on other nodes, asynchronous propagation means the full cross-rank chain is no longer intact.  Once Node A's local volatile copies are reclaimed, this missing data breaks the globally continuous chain required for exact replay. Consequently, the latest recoverable state is severely truncated: in the worst case, recovery falls back to the bare Tier-3 anchor at $V-1$.

Figure~\ref{fig:reclaim}(b) shows that \sysname eliminates this vulnerability by decoupling local creation from deletion. The watermark for version $V$ advances only after its base checkpoint has been durably flushed to Tier-3 on all participating nodes (Nodes A and B). A designated central coordinator (e.g., the process with global distributed rank $0$) then broadcasts this safe watermark, enforcing a unified global boundary for state reclamation across all ranks. Once the watermark advances to version $V$, each rank safely purges obsolete states across all tiers: (i) strictly preceding differential checkpoints located in local, peer, and remote storage; (ii) local base checkpoints older than $V$; and (iii) base checkpoints replicated to peer nodes older than $V$. The base checkpoint at $V$ is preserved as the current globally safe anchor.

%% file: implementation.tex
\section{Implementation}
\label{sec:implementation}

We implement \sysname as a fault-tolerance layer atop PyTorch \cite{paszke19} and DeepSpeed \cite{rasley20} in 12\,K lines of code. \sysname hooks directly into existing optimizer, serialization, RPC, and communication interfaces without rewriting the underlying frameworks. High-level coordination logic is written in Python; performance-critical compression and replay primitives are accelerated via custom CUDA extensions.

\para{Saving path.} \sysname generates differential checkpoints by extracting incremental updates directly from rank-local gradient buffers exposed by the training framework. The compression pipeline dynamically selects between dense INT8 quantization \cite{alistarh17qsgd} for small tensors and sampled-threshold sparse compression \cite{yao25} for large ones. All device-side captures, host-device staging, and serialization tasks are offloaded to low-priority CUDA streams and pinned host memory buffers for fully asynchronous execution. For base checkpoints, \sysname intercepts the internal serialization routine (\texttt{torch.save}) to capture each emitted shard as a zero-copy in-memory byte payload, which is then simultaneously dispatched to Tier-1 local memory, replicated to Tier-2 peer memory, and queued for asynchronous migration to Tier-3. Tier-2 replication uses isolated communication groups that decouple the RPC-based metadata control plane from the tensor-based chunk data plane, and an adaptive chunk-based transmission scheduler that prevents gigabyte-scale base checkpoints from stalling training.

\para{Retrieval path.} Upon restart, each rank independently identify its newest recoverable base checkpoint across surviving tiers via global consensus, then 
materializes only the shards required by its active DP/TP/PP/ZeRO placement following the cost-ordered cascade (i.e., prioritizing Tier-1, then Tier-2 peer replicas, and accessing Tier-3 as a last resort). Differential checkpoints undergo an identical consensus process to establish the safe replay boundary. Compressed updates are reconstructed directly on the recovery GPUs: sparse payloads are fused into dense tensors and INT8 payloads are dequantized from their stored scales. The fused multi-step replay operator consolidates the first $N-1$ differential steps in a replay batch into a single device-side update pass before re-exposing the final step to the native optimizer path, which is necessary to correctly trigger the post-update communication and optimizer state transitions (e.g., gradient synchronization and loss-scaler updates) that the fused kernel does not replicate.



\para{Tier-3 commit and reclamation.} Tier-3 persistence is offloaded to a dedicated background daemon. As remote object stores lack atomic directory operations, the daemon employs a two-phase marker-based commit protocol: each rank uploads its shards and writes a local completion marker; the designated coordinator (e.g., rank 0) then verifies all markers before publishing a global \texttt{\_COMMITTED} signal. Any checkpoint lacking this signal is ignored during recovery. Once committed, the coordinator advances the global watermark, broadcasting it to trigger asynchronous garbage collection on all ranks. The cleanup routine purges: (i) differential checkpoints whose replay range precedes the watermark from Tier-1 and Tier-2, and (ii) base checkpoints older than the watermark from both local and peer storage. Differential checkpoints in Tier-3 are retained for one additional base checkpoint interval to ensure freshly provisioned replacement ranks can always reconstruct state from Tier-3 alone.

%% file: evaluation.tex
\section{Evaluation}
\label{sec:evaluation}

We evaluate \sysname and summarize our findings below:
\begin{itemize}[leftmargin=*]
\item 
{\bf Training efficiency.} \sysname achieves lower training overhead and higher checkpointing frequencies than state-of-the-art checkpointing systems (Exp\#1-3).
\item 
{\bf Recovery efficiency.} \sysname reduces recovery time for heterogeneous failures by leveraging Tiers-1/2's checkpoints, and provides robust fallback to Tier-3 for rack failures (Exp\#4).
\item 
{\bf Scalability and compatibility.} \sysname maintains its performance advantages across different model sizes (Exp\#5) and varying parallelism configurations (Exp\#6).
\item 
{\bf Microbenchmarks.} \sysname's fused multi-step replay accelerates differential checkpoint recovery (Exp\#7); its model convergence is preserved without accuracy degradation (Exp\#8); and its multi-tier global reclamation strictly bounds the storage footprint without risking data loss (Exp\#9).
\end{itemize}

\subsection{Experimental Setup}
\label{subsec:eval_setup}

\para{Testbed environment.} We conduct evaluations on a 16-GPU testbed
provisioned from a distributed cloud platform. The testbed comprises four
compute nodes, each equipped with four NVIDIA A800 (80\,GiB) GPUs and 512\,GiB
of CPU memory, interconnected via a 200\,Gbps InfiniBand network. The remote
Tier-3 storage is backed by NVMe SSD-based Lustre (v2.14.0) over an aggregated
400\,Gbps Ethernet network. The software stack includes Ubuntu 22.04, PyTorch
v2.6.0, DeepSpeed v0.16.4, CUDA 12.4, and NCCL v2.21.5.

\para{Models and workloads.} We evaluate \sysname using representative LLM
architectures by adjusting hidden/intermediate sizes, attention heads, and
layers \cite{wang23,ming26}, as shown in Table~\ref{tab:models}. We train
BERT on SQuAD \cite{rajpurkar18} and other models on WikiText-103 or WikiText-2
\cite{merity16} using the Adam optimizer \cite{kingma15}. While WikiText-2 is
used solely to verify the convergence of the GPT2 124M model (Exp\#8), all other
evaluations employ billion-scale models (up to 40B parameters). Packing
these 40B-parameter models into just 16 GPUs serves as a stress test
that maximizes per-node memory, network, and storage I/O pressure. A system
resilient to this extreme contention will naturally scale to larger clusters,
where the per-GPU checkpoint payload is diluted.

\begin{table}[t]
\setlength{\tabcolsep}{1pt} 
\centering
\small
\caption{Model configurations used in the evaluation. \#AH is the number of attention heads. \#Layers is the number of layers.}
\label{tab:models}
\vspace{-9pt}
\begin{tabular}{l|c|c|c|c}
\hline
\textbf{Model} & \textbf{Hidden Size} & \textbf{Intermediate} &  \textbf{\#AH} & \textbf{\#Layers} \\
\hline
\hline
GPT2 124M & 768 & 3072 & 12 & 12 \\
\hline
10B (GPT2) & 2048 & 10240 & 40 & 46 \\
\hline
20B (GPT2, BERT) & 5120 & 20480 & 40 & 64 \\
\hline
\makecell[l]{40B (GPT2, BERT, \\ RoBERTa, BLOOM)} & 5120 & 20480 & 40 & 128 \\
\hline
\end{tabular}
\vspace{-6pt}
\end{table}

\para{Baselines.} We compare \sysname against three checkpointing baselines: CheckFreq \cite{mohan21}, Gemini \cite{wang23}, and DataStates-LLM \cite{maurya24}, and against the no-checkpoint baseline DeepSpeed ZeRO-3 \cite{rasley20,rajbhandari20zero}. Since Gemini is not open-sourced, we have re-implemented it to the best of our ability based on its original publication. We also extended the original CheckFreq to support ZeRO-3 for fair comparison. 

\para{Default setups.} By default, \sysname captures gradients at every training iteration to construct differential checkpoints. For adaptive gradient compression, we set the small-tensor quantization threshold to $100$K elements and the large-tensor compression ratio to $k=0.01$. The Tier-2 base checkpoint replication uses a maximum chunk size of 256\,MiB to avoid foreground interference. The batching length for saving differential checkpoints is set to $N=5$ \cite{yao25}. The base checkpoint interval is fixed at 50 iterations, except in Exp\#8 where it is extended to 100 iterations to test the replay efficiency of differential checkpoints.  We enable ZeRO-3 and evaluate the compatibility of supporting different parallelisms in Exp\#6.

\para{Failure injection.} We use failure injection in Exp\#4 and Exp\#8. For
Exp\#4, we emulate three recovery scenarios. For software failures, we terminate
the training processes while keeping all nodes alive, and then relaunch the job
on the same cluster. For node failures, we assume any one of the four nodes
fails, replace it with a fresh node, and restart training from the surviving
checkpoints. For rack failures, we conservatively assume that all four nodes in
our 16-GPU testbed reside in the same rack and experience a correlated power
outage; after power is restored, all four nodes are rebooted and training
resumes from the recovered checkpoint state. For Exp\#8, to emulate frequent
interruptions during long-running training, we periodically inject software
failures by terminating the training processes every $\sim$3,000 iterations and
relaunching the job on the same cluster.



\begin{figure}[t]
\centering
\includegraphics[width=3.2in]{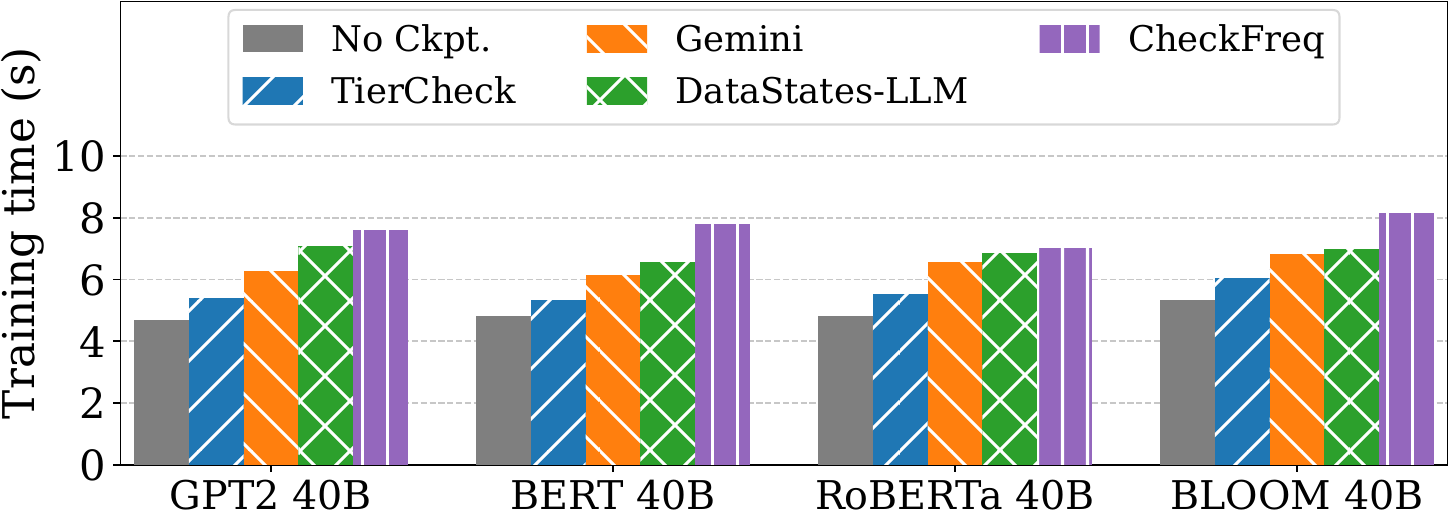} 
\vspace{-9pt}
\caption{(Exp\#1) Average training time per iteration.}
\label{fig:train_time}
\vspace{-6pt}
\end{figure}

\subsection{Macrobenchmarks}
\label{subsec:macro}

\para{(Exp\#1) Average training time per iteration.} We evaluate the average training time over 50 iterations using 40B models with per-iteration snapshots.  Figure~\ref{fig:train_time} shows that \sysname incurs only a 10.7-15.1\% overhead, outperforming state-of-the-art baselines by 52.3-60.8\% (Gemini), 56.9-70.4\% (DataStates-LLM), and 68.2-82.7\% (CheckFreq).

These performance gaps stem from fundamental differences in checkpointing architectures. CheckFreq incurs the highest overhead because it snapshots unpartitioned training states, causing severe synchronous stalls. While DataStates-LLM divides checkpoints into three large partitions, these remain difficult to overlap efficiently and inevitably stall forward passes~\cite{maurya24,ming26}. Gemini mitigates this by fragmenting states into smaller 32\,MiB chunks to fit memory copies within the training pipeline's idle timespans. \sysname bypasses these heavy memory transfers entirely, minimizing foreground overhead by directly intercepting and adaptively compressing native gradients (\S\ref{subsec:save}).


\begin{figure}[t]
\centering
\includegraphics[width=0.95\linewidth]{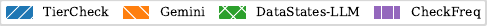}
\begin{minipage}[t]{0.49\linewidth}
\centering
\includegraphics[width=0.95\linewidth]{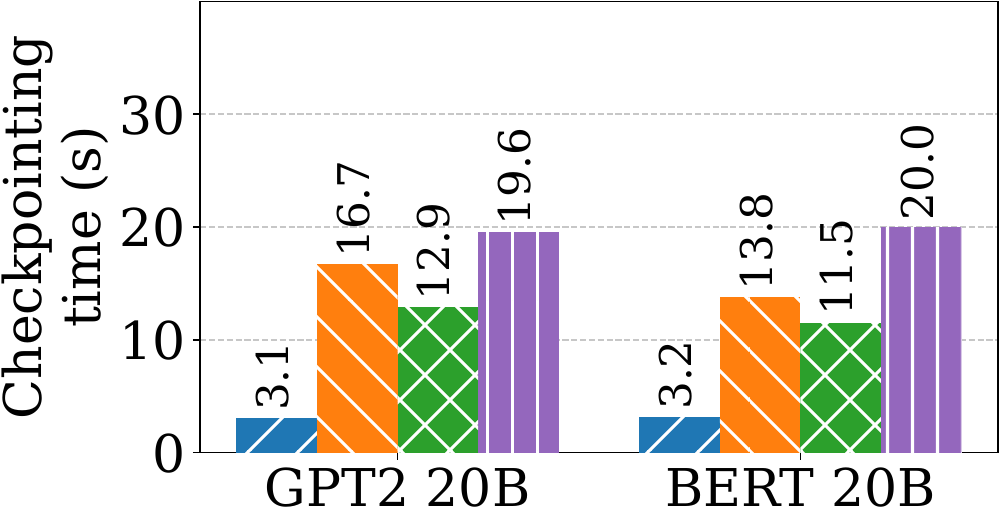} 
\vspace{-9pt}
\caption{(Exp\#2) Checkpointing time.}
\label{fig:ckpt_time}
\vspace{-6pt}
\end{minipage}
\hfill
\begin{minipage}[t]{0.49\linewidth}
\centering
\includegraphics[width=0.95\linewidth]{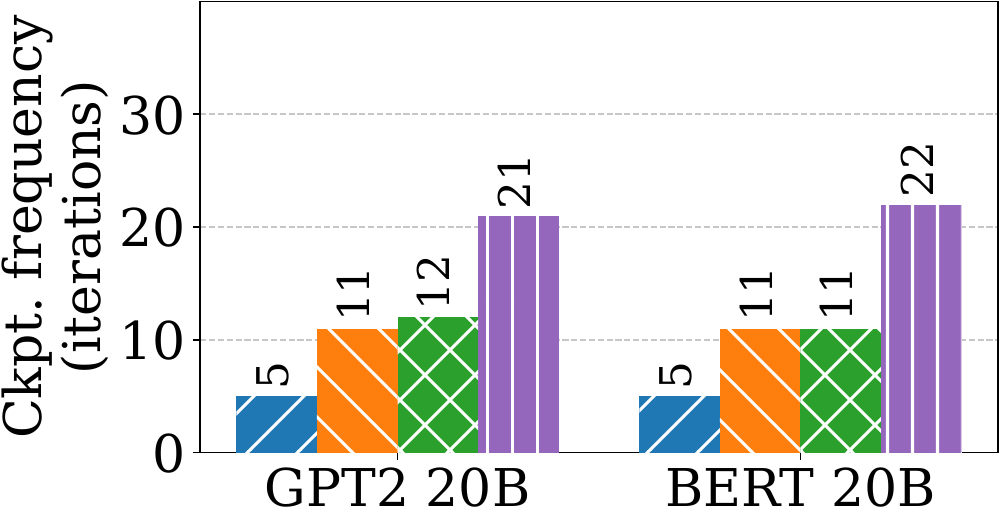} 
\vspace{-9pt}
\caption{(Exp\#3) Checkpoint frequency.}
\label{fig:ckpt_freq}
\vspace{-6pt}
\end{minipage}
\end{figure}

\para{(Exp\#2) Checkpointing time.} We define checkpointing time as the duration required to persist training states to storage tiers. We conduct evaluation on 20B models to prevent prohibitive storage demands. Figure~\ref{fig:ckpt_time} shows that \sysname completes checkpointing in 3.1-3.2\,s per iteration on average for two 20B models. Existing baselines exhibit higher persistence latencies: DataStates-LLM requires 11.5-12.9\,s, Gemini takes 13.8-16.7\,s, and CheckFreq incurs the highest latency at 19.6-20.0\,s per checkpoint. These performance gaps stem from differences in underlying persistence mechanisms. CheckFreq flushes unpartitioned full model states directly to disk and incurs heavy synchronous I/O overhead. Gemini uses small partitions to hide foreground interference, but leaves backend disk flushing unoptimized, resulting in a persistence bottleneck. DataStates-LLM implements asynchronous I/O optimizations in C++. 
In contrast, \sysname persists differential checkpoints in an optimized manner rather than persisting massive full states frequently (\S\ref{subsec:save}).


\para{(Exp\#3) Checkpoint frequency.} We evaluate the maximum checkpointing frequency achievable by each method under a 3.5\% training speed degradation bound~\cite{mohan21,ming26}. We report this frequency as the minimum required checkpoint interval (i.e., the number of training iterations between consecutive saves); a smaller interval signifies a higher checkpointing frequency and thus stronger resilience to failures. Figure~\ref{fig:ckpt_freq} shows that \sysname achieves the highest frequency, requiring an interval of 5 iterations across the evaluated 20B models, which is also consistent to the batching length $N=5$. In contrast, existing methods demand longer intervals to satisfy the same degradation bound: Gemini requires 11 iterations, DataStates-LLM requires 11-12 iterations, and CheckFreq requires 21-22 iterations.


\begin{table}[t]
\setlength{\tabcolsep}{3pt} 
\centering
\small
\caption{(Exp\#4) Recovery time under different failure scenarios.}
\label{tab:recovery_latency}
\vspace{-9pt}
\resizebox{\columnwidth}{!}{ 
\begin{tabular}{l|l|c|c|c}
\hline
\textbf{Systems} & \textbf{Failures} & \textbf{$T_{rollback}$} & \textbf{$T_{rerun}$} & \textbf{Total} \\
\hline
\hline
\multicolumn{5}{c}{\textbf{Model: GPT2 20B}} \\
\hline
CheckFreq & All Scenarios & 17.8\,s & 42.9\,s & 60.5\,s \\
Gemini & All Scenarios & 15.4\,s & 17.9\,s & 33.3\,s \\
DataStates-LLM & All Scenarios & 16.0\,s & 19.9\,s & 35.9\,s \\
\hline
 & Software failures & 4.5\,s & 4.3\,s & 8.8\,s \\
\sysname & Node failures & 9.6\,s & 8.8\,s & 18.4\,s \\
 & Rack failures & 18.4\,s & 5.5\,s & 23.9\,s \\
\hline
\hline
\multicolumn{5}{c}{\textbf{Model: BERT 20B}} \\
\hline
CheckFreq & All Scenarios & 17.8\,s & 45.8\,s & 63.6\,s \\
Gemini & All Scenarios & 16.4\,s & 17.8\,s & 34.2\,s\\
DataStates-LLM & All Scenarios & 16.4\,s & 18.0\,s & 34.4\,s\\
\hline
 & Software failures & 5.5\,s & 4.1\,s & 9.6\,s \\
\sysname & Node failures & 9.8\,s & 8.3\,s & 18.0\,s \\
 & Rack failures & 19.8\,s & 5.9\,s & 25.7\,s \\
\hline
\end{tabular}
}
\vspace{-6pt}
\end{table}

\para{(Exp\#4) Recovery time under different failure scenarios.} We evaluate recovery time, comprising rollback time ($T_{rollback}$) to load the saved checkpoint and rerun time ($T_{rerun}$) to replay lost iterations. Specifically, $T_{rerun}$ is the average of best-case (zero iterations lost) and worst-case (a full interval of iterations lost) re-execution times based on achievable checkpoint frequencies (Exp\#3). For \sysname, $T_{rollback}$ denotes the time to fetch the base checkpoint, while $T_{rerun}$ includes asynchronously pulling differential checkpoints and replaying lost iterations, with the worst-case replay length bounded by its base checkpoint interval of 50 iterations.

Table~\ref{tab:recovery_latency} shows that \sysname recovers in 8.8-9.6\,s for
software failures and 18.0-18.4\,s for node failures. Baselines are
significantly slower; their $T_{rerun}$ alone exceeds \sysname's total software
recovery time. Gemini flushes its volatile buffers to the shared file
system upon failure, degrading all its recoveries to slow Tier-3 fetches.
\sysname's speedup stems from its tiered architecture. For $T_{rollback}$,
\sysname rapidly fetches the base checkpoint from Tier-1 local memory
(4.5-5.5\,s) for software failures and Tier-2 peer memory (9.6-9.8\,s) for node
failures, only incurring higher latencies (18.4-19.8\,s) during rack failures
due to sequential probing before falling back to Tier-3. For $T_{rerun}$,
lightweight differential checkpoints reduce replay overhead to 4.1-8.8\,s for
software failures (\S\ref{subsec:save}). $T_{rerun}$ is slightly higher for node
failures than rack failures (8.3-8.8\,s vs. 5.5-5.9\,s) because the former
relies on the 200\,Gbps training network while the latter fetches directly over
the 400\,Gbps Tier-3 path. Thus, \sysname's localized recovery advantages would
be even more pronounced in deployments with constrained remote bandwidth.



\begin{figure}[t]
\begin{minipage}[t]{0.54\linewidth}
\centering
\begin{tabular}{c}
\includegraphics[width=0.95\linewidth]{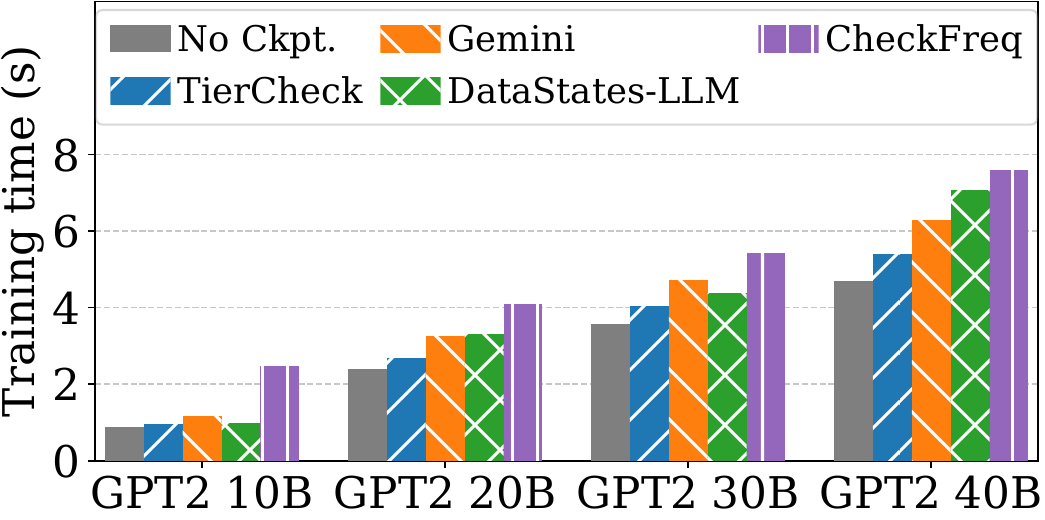} 
\end{tabular}
\vspace{-9pt}
\caption{(Exp\#5) Scalability of different model sizes.}
\label{fig:scala}
\vspace{-6pt}
\end{minipage}
\hfill
\begin{minipage}[t]{0.45\linewidth}
\centering
\begin{tabular}{c}
\includegraphics[width=0.95\linewidth]{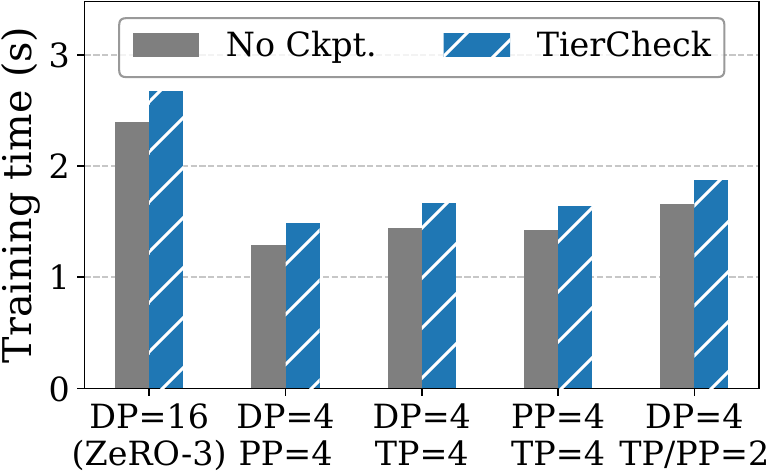} 
\end{tabular}
\vspace{-9pt}
\caption{(Exp\#6) Compatibility with parallelisms.}
\label{fig:parallel}
\vspace{-6pt}
\end{minipage}
\end{figure}

\para{(Exp\#5) Scalability of different model sizes.} We evaluate the scalability of different model sizes by measuring the average training time per iteration with GPT2 models ranging from 10B to 40B parameters. As shown in Figure~\ref{fig:scala}, \sysname consistently outperforms the evaluated baselines across all model scales. While the checkpointing overhead of existing systems grows significantly as the model size increases, \sysname remains closely aligned with the no-checkpoint baseline, incurring a minimal training overhead of only 10.6-15.1\%. This indicates that \sysname's asynchronous transmission and adaptive compression mechanisms maintain consistent performance advantages across varying model dimensions and scale effectively to larger models.

\para{(Exp\#6) Compatibility with parallelisms.} We evaluate five parallelism configurations: DP=16 (ZeRO-3), DP=4/PP=4, DP=4/TP=4, PP=4/TP=4, and balanced 3D parallelism with DP=4/PP=2/TP=2. Since the other checkpointing systems do not support such 3D-parallel training configurations, we compare only against the no-checkpoint DeepSpeed baseline.  Figure~\ref{fig:parallel} shows that \sysname remains consistently close to the no-checkpoint DeepSpeed baseline across all settings, demonstrating its compatibility with both pure data parallelism and hybrid 3D parallelism. The training-time overhead of \sysname over DeepSpeed is 11.8\% for DP=16, 15.8\% for DP/PP=4, 15.2\% for DP/TP=4, 15.3\% for PP/TP=4, and 12.6\% for DP=4/PP=2/TP=2.  Overall, these results indicate that \sysname does not rely on a particular parallelism strategy and maintains low additional overhead across diverse 3D-parallel deployments.

\begin{figure*}[t]
\begin{minipage}[t]{0.33\linewidth}
\centering
\begin{tabular}{c}
\includegraphics[width=0.95\linewidth]{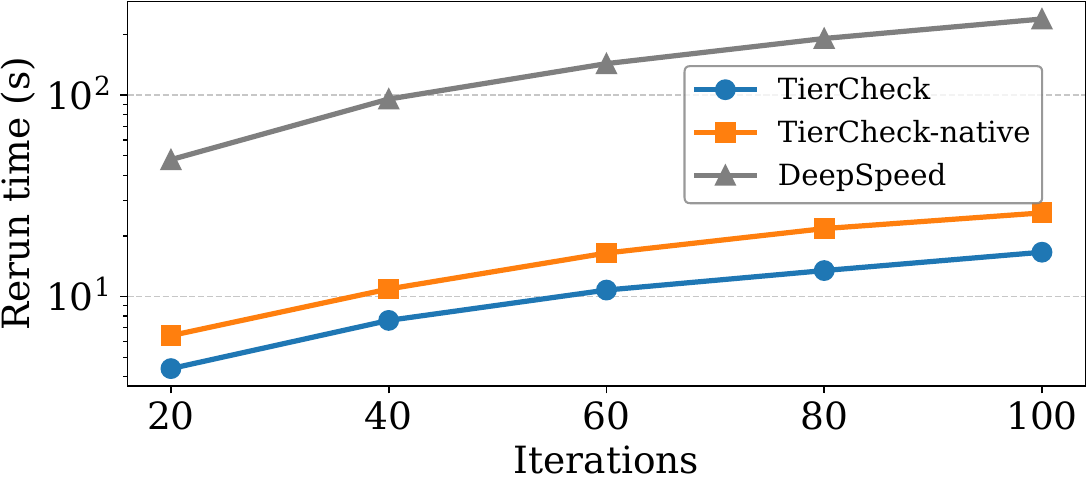}
\end{tabular}
\vspace{-9pt}
\caption{(Exp\#7) Fused multi-step differential checkpoint replay.}
\label{fig:exp_replay}
\vspace{-6pt}
\end{minipage}
\hfill
\begin{minipage}[t]{0.33\linewidth}
\centering
\begin{tabular}{c}
\includegraphics[width=0.95\linewidth]{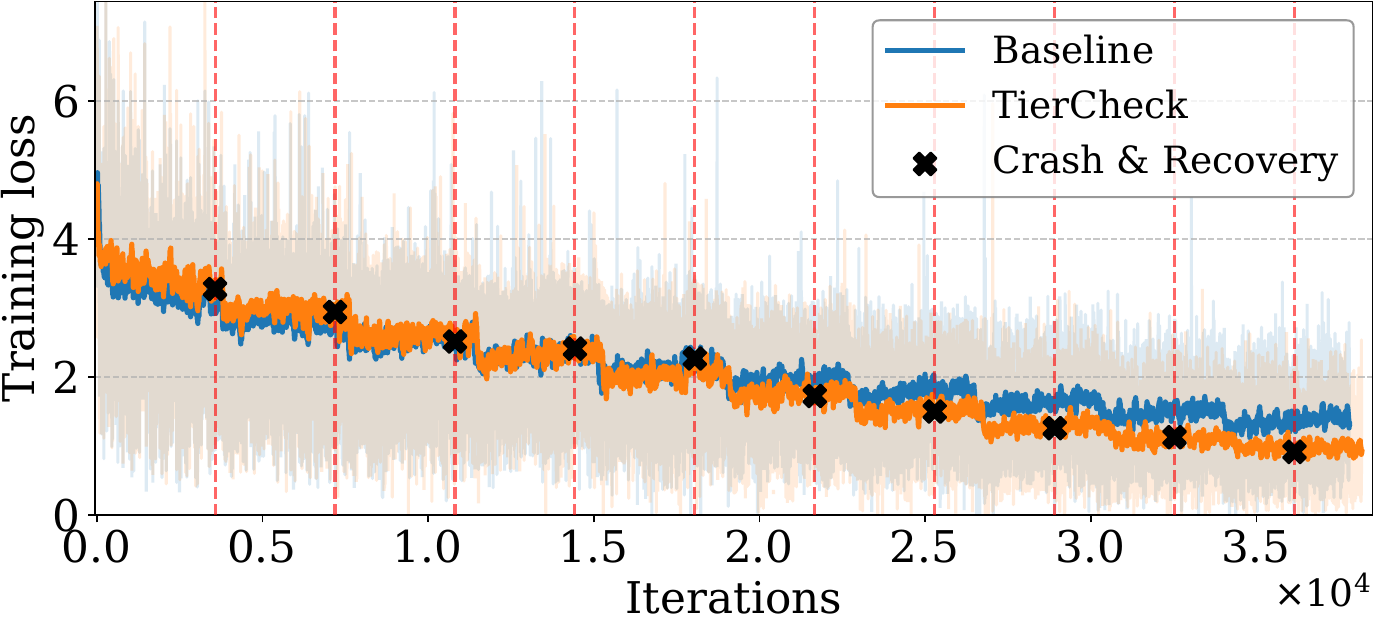}
\end{tabular}
\vspace{-9pt}
\caption{(Exp\#8) Convergence accuracy.}
\label{fig:convergence}
\end{minipage}
\hfill
\begin{minipage}[t]{0.33\linewidth}
\centering
\begin{tabular}{c}
\includegraphics[width=0.95\linewidth]{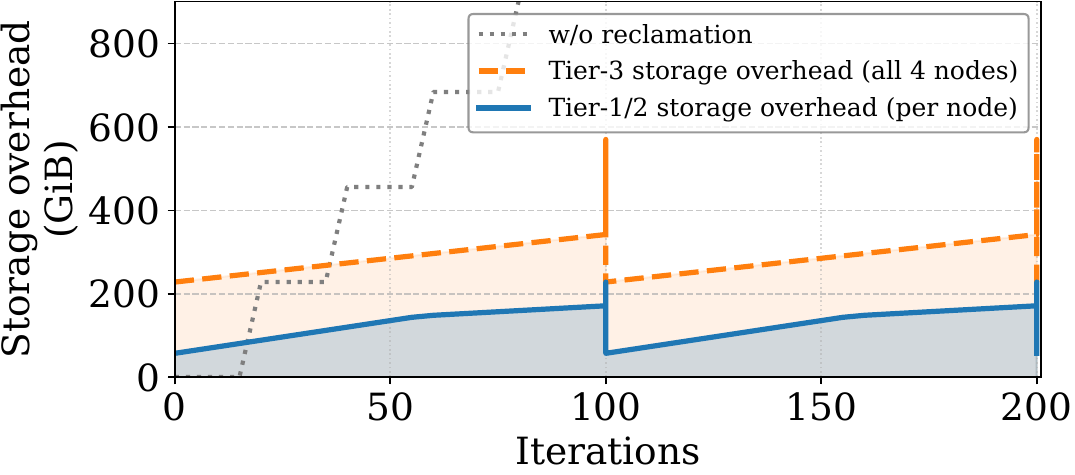}
\end{tabular}
\vspace{-9pt}
\caption{(Exp\#9) Storage overhead.}
\label{fig:storage}
\end{minipage}
\end{figure*}

\subsection{Microbenchmarks}
\label{subsec:micro}

\para{(Exp\#7) Fused multi-step differential checkpoint replay.} Figure~\ref{fig:exp_replay} evaluates the recovery time on a GPT2 20B model, comparing \sysname's fused multi-step replay against sequential differential checkpoint replay and standard DeepSpeed recovery (i.e., rerunning lost iterations). \sysname's fused multi-step replay consistently outperforms the sequential baseline across all evaluated iteration scales. For a recovery chain of 100 iterations, \sysname reduces the rerun time from 26.0\,s to 16.6\,s (i.e., 36.2\% reduction) over sequential replay.  In contrast, DeepSpeed's recovery spends 239.3\,s to process the same interval, making it 14.4$\times$ slower than \sysname. This performance gain is attributed to the fused operator's ability to consolidate multiple optimizer updates into a single pass, thereby mitigating redundant memory I/O and cross-rank communication synchronizations.


\para{(Exp\#8) Convergence accuracy.} To verify algorithmic correctness without
the prohibitive cost of 40B-scale end-to-end training, we fine-tune a GPT2
(124M) model on WikiText-2. This model size features a diverse tensor
distribution that simultaneously triggers INT8 quantization (e.g., for small
biases) and Top-K sparsification (e.g., for large weights). By injecting
failures every $\sim$3,000 iterations, Figure~\ref{fig:convergence} shows
\sysname closely tracks the fault-free baseline and even achieves a slightly
lower final loss. This improvement likely occurs because the TopK compression of
differential checkpoints naturally filters out insignificant gradient
fluctuations \cite{aji17}. Crucially, because larger models contain a vast
number of redundant parameters and naturally exhibit even stronger gradient
sparsity (i.e., more gradients are close to zero) \cite{yao25,liu26}, this
algorithmic robustness effectively scales to 40B configurations without accuracy
loss.

\para{(Exp\#9) Storage overhead.} Figure~\ref{fig:storage} shows a traditional baseline linearly accumulating states to 1,140\,GiB by iteration 100. Conversely, \sysname strictly bounds storage via watermark-driven global reclamation. For a 20B model on 16 GPUs, the steady-state footprint peaks at 342.1\,GiB for global Tier-3 and 171.1\,GiB for per-node Tier-1/2. Because a newly generated base checkpoint momentarily coexists with the previous base checkpoint and accumulated differential checkpoints, the system incurs brief transient spikes (570.1\,GiB in Tier-3, 228.1\,GiB in Tier-1/2) during handoff. Post-reclamation, these footprints rapidly reset to 228\,GiB and 57\,GiB, respectively. This demonstrates that \sysname prevents continuous storage exhaustion, ensuring sustainable training under constrained resources.

%% file: related.tex
\section{Related Work}
\label{sec:relatd}


\para{Optimizing checkpointing efficiency.} 
Prior work buffers training states in local RAM-disks or GPU multi-level caches \cite{xu24,maurya23gpu}, or exploits persistent memory for zero-copy checkpointing \cite{li24}. To overlap I/O with computation, CheckFreq \cite{mohan21}, DataStates-LLM \cite{maurya24}, and PCcheck \cite{strati25} pipeline partitioned checkpoints into host CPU buffers. PyTorch Distributed Checkpoint \cite{pytorch_dcp} provides standardized sharded saving for massive models. To address multi-dimensional parallelism, ByteCheckpoint \cite{wan25} introduces a unified representation to decouple checkpointing from specific parallelism modes, while Universal Checkpointing \cite{lian25} enables seamless resumption under reconfigurable parallelism. At the network level, FlowCheck \cite{huang25} extracts gradients from mirrored traffic using dedicated CPU nodes, and AsymCheck \cite{ming26} employs asymmetric partitioned checkpointing to exploit varying communication idleness across training phases.  While these approaches reduce training overhead or enhance checkpoint flexibility, they share a common limitation: failure recovery remains bottlenecked by slow remote storage I/O, since all persistent state ultimately resides in a single remote tier.  \sysname bypasses remote storage entirely for the common single-node and GPU failures.


\para{Minimizing checkpoint size.} Reducing checkpoint size reduces storage and network I/O pressure. LowDiff \cite{yao25} persists incremental states via compressed gradients; AdaCheck \cite{liu26} adapts to various parallelism modes by profiling tensor redundancy to store only half-precision gradients; MoEvement \cite{gandhi26} introduces sparse checkpointing for Mixture-of-Experts (MoE) models by incrementally snapshotting active expert subsets. However, aggressive size reductions incur recovery overhead: LowDiff incurs sequential replay overhead over accumulated differences; AdaCheck imposes runtime redundancy-detection costs; MoEvement requires localized recomputation for unsnapshotted states.  \sysname applies lightweight Top-K threshold estimation to keep differential payloads small without expensive profiling, and uses fused multi-step operators to accelerate sequential replay.

\para{Fast recovery.} In-memory architectures like Gemini \cite{wang23} and Transom \cite{wu23} preserve snapshots in host CPU memory to achieve near-zero recovery for isolated failures. Swift \cite{zhong24} leverages data-parallelism group replicas combined with selective recomputation, while FT-HSDP \cite{salpekar26} implements replica-level fault tolerance. CheckFree \cite{blagoev26} explores checkpoint-free recovery via weighted averaging of neighboring pipeline stages. MoEvement \cite{gandhi26} enables localized recovery for MoE models via upstream activation logging. \sysname targets heterogeneous failures and maintains fast recovery via cluster-aware tiered checkpointing.

%% file: conclusion.tex
\section{Conclusion}
\label{sec:conclusion}

We present \sysname, a cluster-aware tiered checkpointing system that tolerates
heterogeneous failures in large-scale LLM training. By aligning checkpoint
placement with failure domains, \sysname decouples state persistence into a
high-frequency differential checkpoint stream and a low-frequency base
checkpoint stream, distributing them across local memory, peer memory, and
remote persistent storage. To minimize training stalls, it introduces
adaptive gradient compression and asymmetric transmission scheduling. 
During recovery, it leverages decentralized consensus and fused multi-step
differential checkpoint replaying to enable fast, localized restoration.
Furthermore, a watermark-driven global reclamation mechanism ensures strict
cross-tier consistency and reliably bounds storage footprints. Evaluations on up
to 40B-parameter models demonstrate that \sysname effectively reduces both
checkpointing overhead and recovery time against state-of-the-art baselines.


%% file: paper.bib
@Article{achiam23,
  Title                    = {{GPT}-4 technical report},
  Author                   = {Achiam, Josh and Adler, Steven and Agarwal, Sandhini and Ahmad, Lama and Akkaya, Ilge and Aleman, Florencia Leoni and Almeida, Diogo and Altenschmidt, Janko and Altman, Sam and Anadkat, Shyamal and others},
  Journal                  = {arXiv},
  Year                     = {2023},
  Pages                    = {arXiv preprint arXiv:2303.08774}
}

@InProceedings{aji17,
  Title                    = {Sparse Communication for Distributed Gradient Descent},
  Author                   = {Aji, Alham Fikri and Heafield, Kenneth},
  Booktitle                = {Proc. of EMNLP},
  Year                     = {2017}
}

@InProceedings{alistarh17qsgd,
  Title                    = {{QSGD}: Communication-Efficient {SGD} via Gradient Quantization and Encoding},
  Author                   = {Alistarh, Dan and Grubic, Demjan and Li, Jerry and Tomioka, Ryota and Vojnovic, Milan},
  Booktitle                = {Proc. of NeurIPS},
  Year                     = {2017}
}

@InProceedings{blagoev26,
  Title                    = {All is Not Lost: {LLM} Recovery without Checkpoints},
  Author                   = {Nikolay Blagoev and O\u{g}uzhan Ersoy and Lydia Yiyu Chen},
  Booktitle                = {Proc. of EuroMLSys},
  Year                     = {2026}
}

@InProceedings{bornholt21,
  Title                    = {Using lightweight formal methods to validate a key-value storage node in {Amazon S3}},
  Author                   = {James Bornholt and Rajeev Joshi and Vytautas Astrauskas and Brendan Cully and Bernhard Kragl and Seth Markle and Kyle Sauri and Drew Schleit and Grant Slatton and Serdar Tasiran and Jacob Van Geffen and Andrew Warfield},
  Booktitle                = {Proc. of ACM SOSP},
  Year                     = {2021}
}

@Misc{dean09,
  Title                    = {Designs, lessons and advice from building large distributed systems},

  Author                   = {J. Dean},
  Note                     = {Keynote talk at LADIS},
  Year                     = {2009}
}

@Article{elnozahy02,
  Title                    = {A survey of rollback-recovery protocols in message-passing systems},
  Author                   = {Elnozahy, Elmootazbellah Nabil and Alvisi, Lorenzo and Wang, Yi-Min and Johnson, David B},
  Journal                  = {ACM Computing Surveys},
  Year                     = {2002},
  Number                   = {3},
  Pages                    = {375--408},
  Volume                   = {34}
}

@InProceedings{ford10,
  Title                    = {Availability in Globally Distributed Storage Systems},
  Author                   = {Ford, Daniel and Labelle, Fran{\c{c}}ois and Popovici, Florentina I and Stokely, Murray and Truong, Van-Anh and Barroso, Luiz and Grimes, Carrie and Quinlan, Sean},
  Booktitle                = {Proc. of USENIX OSDI},
  Year                     = {2010}
}

@InProceedings{gandhi26,
  Title                    = {Sparse Checkpointing for Fast and Reliable {MoE} Training},
  Author                   = {Swapnil Gandhi and Christos Kozyrakis},
  Booktitle                = {Proc. of USENIX NSDI},
  Year                     = {2026}
}

@Article{grattafiori24,
  Title                    = {The {Llama} 3 herd of models},
  Author                   = {Grattafiori, Aaron and Dubey, Abhimanyu and Jauhri, Abhinav and Pandey, Abhinav and Kadian, Abhishek and Al-Dahle, Ahmad and Letman, Aiesha and Mathur, Akhil and Schelten, Alan and Vaughan, Alex and others},
  Journal                  = {arXiv},
  Year                     = {2024},
  Pages                    = {arXiv preprint arXiv:2407.21783}
}

@InProceedings{huang19,
  Title                    = {{GPipe}: Efficient Training of Giant Neural Networks using Pipeline Parallelism},
  Author                   = {Huang, Yanping and Cheng, Youlong and Bapna, Ankur and Firat, Orhan and Chen, Dehao and Chen, Mia Xu and Lee, HyoukJoong and Ngiam, Jiquan and Le, Quoc V and Wu, Yonghui and Chen, Zhifeng},
  Booktitle                = {Proc. of NeurIPS},
  Year                     = {2019}
}

@InProceedings{huang25,
  Title                    = {{FlowCheck}: Decoupling Checkpointing and Training of Large-Scale Models},
  Author                   = {Huang, Zimeng and Nie, Hao and Jia, Haonan and Jiang, Bo and Guo, Junchen and Lu, Jianyuan and Wen, Rong and Lyu, Biao and Zhu, Shunmin and Wang, Xinbing},
  Booktitle                = {Proc. of EuroSys},
  Year                     = {2025}
}

@InProceedings{jiang24,
  Title                    = {{MegaScale}: Scaling large language model training to more than 10,000 {GPUs}},
  Author                   = {Jiang, Ziheng and Lin, Haibin and Zhong, Yinmin and Huang, Qi and Chen, Yangrui and Zhang, Zhi and Peng, Yanghua and Li, Xiang and Xie, Cong and Nong, Shibiao and others},
  Booktitle                = {Proc. of USENIX NSDI},
  Year                     = {2024}
}

@InProceedings{kingma15,
  Title                    = {Adam: A method for stochastic optimization},
  Author                   = {Kingma, Diederik P and Ba, Jimmy},
  Booktitle                = {Proc. of ICLR},
  Year                     = {2015}
}

@Article{scao22bloom,
  Title                    = {{BLOOM}: A {176B}-Parameter Open-Access Multilingual Language Model},
  Author                   = {Le Scao, Teven and Fan, Angela and Akiki, Christopher and Pavlick, Ellie and Ili{\'c}, Suzana and Hesslow, Daniel and Castagn{\'e}, Roman and Luccioni, Alexandra Sasha and Yvon, Fran{\c{c}}ois and Gall{\'e}, Matthias and Tow, Jonathan and Rush, Alexander M. and Biderman, Stella and Webson, Albert and Ammanamanchi, Pawan Sasanka and Wang, Thomas and Sagot, Beno{\^i}t and Muennighoff, Niklas and del Moral, Albert Villanova and Ruwase, Olatunji and Bawden, Rachel and Bekman, Stas and McMillan-Major, Angelina and Wolf, Thomas and Beltagy, Iz and Nguyen, Huu and Saulnier, Lucile and Tan, Samson and Suarez, Pedro Ortiz and Sanh, Victor and Lauren{\c{c}}on, Hugo and Jernite, Yacine and Launay, Julien and Mitchell, Margaret and Raffel, Colin},
  Journal                  = {arXiv preprint arXiv:2211.05100},
  Year                     = {2022}
}

@InProceedings{li24,
  Title                    = {Portus: Efficient {DNN} checkpointing to persistent memory with zero-copy},
  Author                   = {Li, Yuanhao and Wu, Tianyuan and Li, Guancheng and Song, Yanjie and Yin, Shu},
  Booktitle                = {Proc. of ICDCS},
  Year                     = {2024}
}

@InProceedings{lian25,
  Title                    = {Universal Checkpointing: A Flexible and Efficient Distributed Checkpointing System for Large-Scale {DNN} Training with Reconfigurable Parallelism},
  Author                   = {Xinyu Lian and Sam Ade Jacobs and Lev Kurilenko and Masahiro Tanaka and Stas Bekman and Olatunji Ruwase and Minjia Zhang},
  Booktitle                = {Proc. of USENIX ATC},
  Year                     = {2025}
}

@InProceedings{liu26,
  Title                    = {{AdaCheck}: An Adaptive Checkpointing System for Efficient {LLM} Training with Redundancy Utilization},
  Author                   = {Liu, Weijie and Li, Shengwei and Lai, Zhiquan and Ge, Keshi and Chen, Qiaoling and Sun, Peng and Li, Dongsheng and Lu, Kai},
  Booktitle                = {Proc. of USENIX FAST},
  Year                     = {2026}
}

@InProceedings{maurya23gpu,
  Title                    = {{GPU}-enabled asynchronous multi-level checkpoint caching and prefetching},
  Author                   = {Maurya, Avinash and Rafique, M Mustafa and Tonellot, Thierry and AlSalem, Hussain J and Cappello, Franck and Nicolae, Bogdan},
  Booktitle                = {Proc. of HPDC},
  Year                     = {2023}
}

@InProceedings{maurya24,
  Title                    = {DataStates-{LLM}: Lazy asynchronous checkpointing for large language models},
  Author                   = {Maurya, Avinash and Underwood, Robert and Rafique, M Mustafa and Cappello, Franck and Nicolae, Bogdan},
  Booktitle                = {Proc. of HPDC},
  Year                     = {2024}
}

@Article{merity16,
  Title                    = {Pointer sentinel mixture models},
  Author                   = {Merity, Stephen and Xiong, Caiming and Bradbury, James and Socher, Richard},
  Journal                  = {arXiv preprint arXiv:1609.07843},
  Year                     = {2016}
}

@InProceedings{ming26,
  Title                    = {{AsymCheck}: Asymmetric Partitioned Checkpointing for Efficient Large Language Model Training},
  Author                   = {Ming, Zhangqiang and Hu, Yuchong and Luo, Zhiyuan and Lee, Patrick P. C. and Shu, Yuanhao and Zhou, Wenxiang and Feng, Dan},
  Booktitle                = {Proc. of ACM/IEEE DAC},
  Year                     = {2026}
}

@InProceedings{mohan21,
  Title                    = {{CheckFreq}: Frequent, Fine-Grained {DNN} Checkpointing},
  Author                   = {Mohan, Jayashree and Phanishayee, Amar and Chidambaram, Vijay},
  Booktitle                = {Proc. of USENIX FAST},
  Year                     = {2021}
}

@InProceedings{paszke19,
  Title                    = {{PyTorch}: An Imperative Style, High-Performance Deep Learning Library},
  Author                   = {Adam Paszke and Sam Gross and Francisco Massa and Adam Lerer and James Bradbury and Gregory Chanan and Trevor Killeen and Zeming Lin and Natalia Gimelshein and Luca Antiga and Alban Desmaison and Andreas Kopf and Edward Yang and Zachary DeVito and Martin Raison and Alykhan Tejani and Sasank Chilamkurthy and Benoit Steiner and Lu Fang and Junjie Bai and Soumith Chintala},
  Booktitle                = {Proc. of NeurIPS},
  Year                     = {2019}
}

@Misc{pytorch_dcp,
  Title                    = {Distributed Checkpoint ({DCP}) --- {PyTorch} Tutorials},

  Author                   = {{PyTorch Team}},
  HowPublished             = {\url{https://docs.pytorch.org/tutorials/recipes/distributed_checkpoint_recipe.html}},
  Year                     = {2024}
}

@InProceedings{rajbhandari20zero,
  Title                    = {{ZeRO}: Memory optimizations toward training trillion parameter models},
  Author                   = {Rajbhandari, Samyam and Rasley, Jeff and Ruwase, Olatunji and He, Yuxiong},
  Booktitle                = {Proc. of SC},
  Year                     = {2020}
}

@Article{rajpurkar18,
  Title                    = {Know what you don't know: Unanswerable questions for {SQuAD}},
  Author                   = {Rajpurkar, Pranav and Jia, Robin and Liang, Percy},
  Journal                  = {arXiv preprint arXiv:1806.03822},
  Year                     = {2018}
}

@InProceedings{rasley20,
  Title                    = {{DeepSpeed}: System optimizations enable training deep learning models with over 100 billion parameters},
  Author                   = {Rasley, Jeff and Rajbhandari, Samyam and Ruwase, Olatunji and He, Yuxiong},
  Booktitle                = {Proc. of KDD},
  Year                     = {2020}
}

@InProceedings{renggli19,
  Title                    = {{SparCML}: High-Performance Sparse Communication for Machine Learning},
  Author                   = {Renggli, Cedric and Ashkboos, Saleh and Aghagolzadeh, Mehdi and Alistarh, Dan and Hoefler, Torsten},
  Booktitle                = {Proceedings of the International Conference for High Performance Computing, Networking, Storage and Analysis (SC)},
  Year                     = {2019},
  Publisher                = {ACM}
}

@Article{salpekar26,
  Title                    = {Training {LLMs} with Fault Tolerant {HSDP} on 100,000 {GPUs}},
  Author                   = {Omkar Salpekar and Rohan Varma and Kenny Yu and Vladimir Ivanov and Yang Wang and Ahmed Sharif and Min Si and Shawn Xu and Feng Tian and Shengbao Zheng and others},
  Journal                  = {arXiv preprint arXiv:2602.00277},
  Year                     = {2026}
}

@InProceedings{schwan03,
  Title                    = {Lustre: Building a file system for 1,000-node clusters},
  Author                   = {Philip Schwan},
  Booktitle                = {Proc. of Linux Symposium},
  Year                     = {2003}
}

@Article{shoeybi19,
  Title                    = {Megatron-{LM}: Training multi-billion parameter language models using model parallelism},
  Author                   = {Shoeybi, Mohammad and Patwary, Mostofa and Puri, Raul and LeGresley, Patrick and Casper, Jared and Catanzaro, Bryan},
  Journal                  = {arXiv},
  Year                     = {2019},
  Pages                    = {arXiv preprint arXiv:1909.08053}
}

@InProceedings{strati25,
  Title                    = {{PCcheck}: Persistent Concurrent Checkpointing for {ML}},
  Author                   = {Strati, Foteini and Friedman, Michal and Klimovic, Ana},
  Booktitle                = {Proc. of ACM ASPLOS},
  Year                     = {2025}
}

@InProceedings{wan25,
  Title                    = {{ByteCheckpoint}: A Unified Checkpointing System for Large Foundation Model Development},
  Author                   = {Borui Wan and Mingji Han and Yiyao Sheng and Yanghua Peng and Haibin Lin and Mofan Zhang and Zhichao Lai and Menghan Yu and Junda Zhang and Zuquan Song and Xin Liu and Chuan Wu},
  Booktitle                = {Proc. of USENIX NSDI},
  Year                     = {2025}
}

@InProceedings{wang23,
  Title                    = {Gemini: Fast failure recovery in distributed training with in-memory checkpoints},
  Author                   = {Wang, Zhuang and Jia, Zhen and Zheng, Shuai and Zhang, Zhen and Fu, Xinwei and Ng, TS Eugene and Wang, Yida},
  Booktitle                = {Proc. of SOSP},
  Year                     = {2023}
}

@InProceedings{weil06,
  Title                    = {{Ceph}: A scalable, high-performance distributed file system},
  Author                   = {Weil, Sage A and Brandt, Scott A and Miller, Ethan L and Long, Darrell DE and Maltzahn, Carlos},
  Booktitle                = {Proc. of USENIX OSDI},
  Year                     = {2006}
}

@Article{wu23,
  Title                    = {Transom: An efficient fault-tolerant system for training {LLMs}},
  Author                   = {Wu, Baodong and Xia, Lei and Li, Qingping and Li, Kangyu and Chen, Xu and Guo, Yongqiang and Xiang, Tieyao and Chen, Yuheng and Li, Shigang},
  Journal                  = {arXiv},
  Year                     = {2023},
  Pages                    = {arXiv preprint arXiv:2310.10046}
}

@InProceedings{xu24,
  Title                    = {An Efficient Checkpointing System for Large Machine Learning Model Training},
  Author                   = {Xu, Wubiao and Huang, Xin and Meng, Shiman and Zhang, Weiping and Guo, Luanzheng and Sato, Kento},
  Booktitle                = {Proc. of SC Workshops},
  Year                     = {2024}
}

@InProceedings{yao25,
  Title                    = {{LowDiff}: Efficient Frequent Checkpointing via Low-Cost Differential for High-Performance Distributed Training Systems},
  Author                   = {Yao, Chenxuan and Hu, Yuchong and Liu, Feifan and Liu, Zhengyu and Feng, Dan},
  Booktitle                = {Proc. of SC},
  Year                     = {2025}
}

@Article{zhang19,
  Title                    = {{SimEDC}: A Simulator for the Reliability Analysis of Erasure-Coded Data Centers},
  Author                   = {Mi Zhang and Shujie Han and Patrick P. C. Lee},
  Journal                  = {IEEE Transactions on Parallel and Distributed Systems},
  Year                     = {2019},
  Number                   = {12},
  Pages                    = {2836--2848},
  Volume                   = {30}
}

@InProceedings{zhang20,
  Title                    = {An empirical study on program failures of deep learning jobs},
  Author                   = {Zhang, Ru and Xiao, Wencong and Zhang, Hongyu and Liu, Yu and Lin, Haoxiang and Yang, Mao},
  Booktitle                = {Proc. of ACM/IEEE ICSE},
  Year                     = {2020}
}

@Article{zhong24,
  Title                    = {Swift: Expedited Failure Recovery for Large-Scale {DNN} Training},
  Author                   = {Zhong, Yuchen and Sheng, Guangming and Liu, Juncheng and Yuan, Jinhui and Wu, Chuan},
  Journal                  = {IEEE Transactions on Parallel and Distributed Systems},
  Year                     = {2024},
  Number                   = {9},
  Pages                    = {1644--1656},
  Volume                   = {35}
}
